\documentclass[pdflatex,letterpaper]{mn2e}
\pdfoutput=1
\usepackage{verbatim}

\usepackage{aas_macros}
\usepackage{natbib}
\usepackage[pdftex]{graphicx}
\usepackage{amsmath}
\usepackage[english]{babel}
\providecommand{\e}[1]{\ensuremath{\times 10^{#1}}}

\title[The Growth and Evolution of Milky Way-mass Galaxies]{The Diversity of Growth Histories of Milky Way-mass Galaxies}

\author[Terrazas et al. 2016]{Bryan A. Terrazas$^{1}$\thanks{E-mail: bterraza@umich.edu}, Eric F. Bell$^{1}$, Bruno M. B. Henriques$^{2}$, \and Simon D. M. White$^{2}$\\
\\
$^{1}$Department of Astronomy, University of Michigan, 311 West Hall, 1085 South University Ave, Ann Arbor, MI 48109-1107\\
$^{2}$Max-Planck-Institut f{\"u}r Astrophysik, Karl-Schwarzschild-Str. 1, 85741 Garching b. M{\"u}nchen, Germany}

\begin{document}

\date{Accepted 2016 March 18. Received 2016 March 14; in original form 2015 July 14}

\pagerange{\pageref{firstpage}--\pageref{lastpage}} \pubyear{2016}

\maketitle

\label{firstpage}

\begin{abstract}
We use the semi-analytic model developed by \citet{hwt2015} to explore the origin of star formation history diversity for galaxies that lie at the centre of their dark matter haloes and have present-day stellar masses in the range $5-8 \times 10^{10}$ M$_{\odot}$, similar to that of the Milky Way. In this model, quenching is the dominant physical mechanism for introducing scatter in the growth histories of these galaxies. We find that present-day quiescent galaxies have a larger variety of growth histories than star-formers since they underwent `staggered quenching' -- a term describing the correlation between the time of quenching and present-day halo mass. While halo mass correlates broadly with quiescence, we find that quiescence is primarily a function of black hole mass, where galaxies quench when heating from their active galactic nuclei becomes sufficient to offset the redshift-dependent cooling rate. In this model, the emergence of a prominent quiescent population is the main process that flattens the stellar mass--halo mass relation at mass scales at or above that of the Milky Way.
\end{abstract}


\section{Introduction}
\label{sec:Intro}

One of the primary goals of galaxy formation theory is to better understand the evolution of galaxy stellar mass in relation to their dark matter haloes. Tight constraints on the parameters of the $\Lambda$CDM cosmological framework from observations have allowed detailed N-body simulations to characterize the distribution of dark matter at the present day and its evolution with cosmic time \citep{swj2005, bsw2009, ktp2011}. In conjunction, observational surveys of hundreds of thousands of galaxies have also begun to allow a statistical understanding of observed galaxy properties over about 10 Gyrs (e.g., SDSS: \citealp{yaa2000}; 2dFGRS: \citealp{cdm2001}; 6dFGRS: \citealp{jsc2004}; GALEX: \citealp{mfs2005}; 2MASS: \citealp{scs2006}; COSMOS: \citealp{sab2007}; CANDELS: \citealp{kff2011, gkf2011}; UltraVISTA: \citealp{mmd2012}, DEEP2: \citealp{ncd2013}). These studies have provided the foundation for understanding the evolution of galaxy number density as a function of stellar mass -- the stellar mass function -- from $z = 0 - 8$ \citep{cnb2001, lw2009, mvf2009, isl2010, mcb2011, iml2013, mms2013, mca2013, tqt2014}. As a result, understanding how the dark matter halo mass function from N-body simulations and the stellar mass functions from observational surveys map onto one another at different epochs has recently been a subject of intense study. 

In this paper, we use the semi-analytic model developed by \citet{hwt2015} to characterize the growth histories of galaxies at the centres of their dark matter haloes with present-day stellar masses $5-8 \times 10^{10}$ M$_{\odot}$ since $z \sim 2.07$. We explore what drives the diversity of pathways that could lead to galaxies with the same stellar mass yet different galactic properties at the present day in the context of this model. Understanding this fundamental behavior of galaxy growth has important implications for the way stellar mass is generally linked with halo mass in a variety of models and observational studies.

A powerful approach for building intuition about how the observable properties of galaxies relate to the dark matter framework is the use of galaxy formation models. The goal of such models is to accurately simulate the gravitationally-driven evolution of dark matter haloes, the infall, cooling, and heating of gas in this dark matter framework, the formation of stars at the centres of potential wells, the formation of rotationally-supported discs and dispersion-supported spheroids, and feedback from stars and black holes. Realistically reproducing these processes is essential to understanding the underlying physics of observed phenomena at extragalactic scales. 

One such approach to simulating galaxy formation, and the approach we choose to use in our study, is semi-analytic modeling \citep{kcd1999, swt2001, hcf2003, hdn2003, kjm2005, csw2006, shc2008, bbm2006, gwb2011, psp2014, hwt2015}. These models use simplified analytic parameterizations to model complex baryonic physics on top of dark matter simulations. This method has the advantage of being relatively computationally inexpensive and therefore more easily able to simulate large cosmological volumes. In addition, searching through parameter space in semi-analytic models is straightforward, especially compared to hydrodynamic simulations. A caveat is that these models incorporate many free parameters, leading to considerable degeneracies in their results. Even so, they have included progressively more nuanced prescriptions for the physical drivers of galaxy evolution, and advanced statistical procedures such as Monte Carlo Markov Chain (MCMC) methods which aim to more comprehensively constrain model parameters using observational data. The development of MCMC methods in semi-analytic models began with the work of \citet{kts2008} and \citet{hto2009} and has since been extended to a wide range of simulations and sampling methods \citep{bb2010, bvg2010, ht2010, lmw2011, lmk2012, hwt2013, mpc2013,  b2014, rcp2015}. As a result, these models have developed into powerful tools to study the mapping of stellar mass onto dark matter haloes.

Models such as these provide tools with which to test and explore physical and statistical recipes for galaxy evolution using large-scale observational datasets. In the past decade, studies attempting to track the ancestry of galaxies by linking galaxy populations in these datasets at different redshifts have flourished \citep{db2007, vwb2010, bwv2011, pff2011, lvf2013, vln2013, bfp2013, pvf2013, bfp2014, wgc2014}. While it is not possible to observe how individual systems evolve, a combination of models and observations can be used to develop methods by which one can attempt to identify the progenitors of present-day galaxies. 

Many of these progenitor studies have used stellar mass to characterize the growth histories of galaxies. Such studies suggest that galaxies grow significantly in size but without much gain in mass for the most massive galaxies \citep{vwb2010, min2015}, whereas lower mass galaxies grow significantly in both mass and size \citep{pff2013, plq2015}. These studies depend on strong assumptions, such as a constant comoving number density or a stellar mass growth inferred from the evolution of the star forming main sequence. Yet, galaxy growth may involve a considerable degree of stochastic variation as a result of many different halo parameters and assembly histories, leading to a diversity of galaxy growth histories \citep{sls2015, hwt2015, cbn2015, twm2015}. A concern is that the degree of growth history diversity may be large enough to undermine any insight gained by studying the average or median growth histories of galaxy populations.

Intuitively, intrinsic scatter in growth histories is a natural consequence of the halting of star formation in galaxies. A relatively massive galaxy that halts its production of stars by $z = 1$ may end up having the same stellar mass at the present day as a low mass galaxy at high redshift that has continually formed stars. Observational surveys have shown that the quiescent population of galaxies has grown substantially since $z \sim 2$ \citep{bwm2004, fww2007, bwv2009, mms2013}. While quenched central galaxies are possible at stellar masses above $10^{9}$ M$_{\odot}$ \citep{gby2012}, they become increasingly more common at high stellar masses \citep{khw2003}. Detailed studies of these quenched galaxies have revealed concentrated light distributions and high velocity dispersions \citep{b2008, fvs2008, cfk2012, bvp2012, lws2014}, pointing to the likely existence of relatively large central black holes. 

For this reason, one of the most popular explanations for the quenching of galaxies at high stellar mass is feedback from active galactic nuclei (AGN) since heating from this mechanism is thought to correlate with black hole mass. This mechanism works first via quasar-mode then radio-mode feedback \citep{kh2000, csw2006, ssd2007}. Quasar-mode feedback occurs as a result of mergers and drives cold gas into the central regions of the galaxy. This causes rapid growth of the black hole and high accretion disc luminosities \citep{kh2000, dsh2005}, as well as massive outflows of gas \citep{sgv2011, cms2014}. After this phase, radio-mode feedback begins where gas from the hot halo is fed into the black hole inefficiently, producing a jet which heats the surrounding gaseous atmosphere \citep{psv2012, gbt2012, dgp2013, mgt2014}.

Inclusion of AGN heating in galaxy formation models has significantly improved the agreement between the high-mass ends of the simulated and observed stellar mass functions by reducing the amount of star formation in high mass galaxies \citep{csw2006, bbm2006}. Other less important quenching mechanisms that affect central galaxies are mergers which could deplete the amount of cold gas available in the galaxy by triggering star formation \citep{mh1996, lta2003, wgb2006, bac2007}, secular processes such as morphological quenching \citep{kk2004, mbt2009, cjb2011}, and halo or mass quenching which ties quenching together with hot gas mass \citep{db2006, cdd2006_2, bdn2007, dbe2009, gd2015}.

As a result, the focus of this paper is twofold: (1) to understand the most important parameters that determine how quenching operates in a particular galaxy formation model, and (2) to understand how this quenching adds scatter to stellar mass growth histories for galaxies with present-day stellar masses similar to that of the Milky Way. We will use the semi-analytic model developed by \citet{hwt2015} in order to obtain galaxy growth histories of Milky Way-mass galaxies. We choose to analyse this model as it matches the stellar mass functions and the star-forming/quiescent fractions out to $z \sim 3$ by design (See their Figures 2 and 5, respectively). Agreement with these two observations is essential since our study focuses on stellar mass buildup within the star-forming and quiescent populations. 

For this study, we define Milky Way-mass galaxies as \textit{central} galaxies with stellar masses, $M_{*}$ = 5--8\e{10} M$_{\odot}$ \citep{fhp2006, mc2011}. In this model, central galaxies are defined as those which are located at the minimum of the potential of the main halo. Milky Way-mass galaxies are ideal for studying the many pathways of galaxy evolution since they contain a large diversity of morphologies and star formation histories \citep[e.g.][]{khw2003, bmb2006, mca2013, tqt2014}. The study of Milky Way-mass galaxies therefore allows us to better understand the physical mechanisms that differentiate those galaxies that become star-forming from those that become quiescent within the same stellar mass range at the present day. We choose to focus on central galaxies since satellites are affected by additional processes such as ram pressure stripping, tidal forces, and a loss of hot gas mass when plunging into the tidal field and diffuse gaseous halo of the main galaxy/group/cluster potential. 

The organization of the paper is as follows. After introducing the \citet{hwt2015} semi-analytic model (Section~\ref{sec:Henriques}), we describe their physically-motivated model parameterization of quiescence (Section~\ref{sec:quenching}). We then highlight general trends observed in the stellar mass-halo mass (SMHM) relation of central galaxies and describe the evolutionary pathways of the main progenitors of Milky Way-mass galaxies in the model since $z = 2.07$ (Section~\ref{sec:MWgrowth}). Splitting the present-day Milky Way-mass galaxy population into star-forming and quiescent galaxies, we then examine sources of scatter in the growth histories of each group (Section~\ref{sec:SFQPopulations}). A discussion of how the growth of the black hole contributes to the scatter in Milky Way-mass growth histories (Section~\ref{sec:BHmassdependence}) then leads to a possible way to relate our analysis to the entire central galaxy population in terms of the SMHM relation (Section~\ref{sec:SMHM_full}). We then focus on how our results may be useful for observational studies by comparing model values of stellar mass, halo mass, and black hole mass as potential observational signatures of the relevant quenching mechanism at work (Section~\ref{sec:observational}). Finally, we discuss some important implications and conclusions from our study on the scatter in the growth histories of Milky Way-mass galaxies (Sections~\ref{sec:discussion} and~\ref{sec:conclusions}).

\section{The Henriques et al. 2015 Semi-Analytic Model}
\label{sec:Henriques}

\subsection{Model Description}
\label{sec:modeldescription}

\citet{hwt2015}, hereafter referred to as H15, produced a semi-analytic model of baryonic processes overlaid on the Millennium Simulation \citep[MS,][]{swj2005}. A second simulation, the Millennium-II Simulation \citep[MS-II,][]{bsw2009}, was run with 125 times better mass resolution, 5 times better force resolution, and 5 times smaller box length than the MS in order to better model the behavior of smaller structures. Combined, the MS and MS-II provide a way to study the formation of galaxies ranging from faint dwarfs to the most massive cD galaxies. The H15 data was downloaded from the Millennium Databases\footnote{To access the Millennium databases: \url{http://gavo.mpa-garching.mpg.de/Millennium}. For a description of the Munich Galaxy Formation Model: \url{http://galformod.mpa-garching.mpg.de/public/LGalaxies}.}.

The H15 model is a descendant of the semi-analytic model produced by \citet{gwb2011} and includes significant improvements in terms of its agreement with observational data. With regards to the dark matter structure, the \citet{gwb2011} model adopts a $\Lambda$CDM cosmology with cosmological parameters based on results from 2dFGRS \citep{cdm2001} and WMAP1 \citep{svp2003}. H15 also adopts a $\Lambda$CDM cosmology but with more recently published cosmological parameters from the \textit{Planck} Collaboration \citep{planck2014}. The new parameters based on \textit{Planck} data are $\Omega_{\rm{M}} = 0.315$, $\Omega_{\Lambda} = 0.685$, $\Omega_{\rm{b}} = 0.0487$ ($f_{\rm{b}} = 0.155$), $n = 0.96$, $\sigma_{\rm{8}} = 0.829$, and $H_{0} = 67.3$ km s$^{-1}$ Mpc$^{-1}$. The Millennium Simulation was scaled to this cosmology according to the technique detailed in \citet{aw2010} and \citet{ah2014}. With this new cosmology, the MS has a resolution of $2160^{3}$ particles in a periodic box of side length $480.279 h^{-1}$ Mpc with a particle mass of $9.6 \e{8} h^{-1}$ M$_{\odot}$. Despite this difference, the change in cosmology does not significantly change the outcome of the model since the uncertainties are dominated by galaxy formation physics rather than cosmology. 

The following description details the general physical mechanisms that affect the evolution of central galaxies since these are the focus of our paper. In this model, there are six main baryonic components that are followed as galaxies evolve in time -- a hot gas atmosphere, cold interstellar gas, a reservoir of gas which has been ejected by winds, stars in the bulge, disc, and intracluster light components, central supermassive black holes, and diffuse primordial gas associated with dark matter that does not yet belong to any halo. These components are functions of the dark matter merger trees on top of which they are built. Primordial gas is fed into the halo in one of two ways: (1) Rapidly infalling at the free-fall time, or (2) forming a cooling flow after the gas has been initially shock heated to the virial temperature \citep{wr1978, wf1991, bd2003}. These two regimes depend on whether the cooling time is shorter or longer than the halo sound crossing time. 

The angular momentum of the gas that cools to the bottom of the potential well leads to the formation of a disc \citep{fe1980}. Once the gas is in the disc, stars can form at a rate that depends on the angular momentum of the disc, the amount of cold gas available, and the maximum circular velocity of the halo. As stars reach the end of their lives, supernovae provide an important source of feedback which can eject gas out of the galaxy and into a reservoir \citep{wr1978, ds1986, ham1990, c1991, wf1991}. This heated gas remains in the reservoir until it is able to join the hot halo and possibly cool back onto the central galaxy depending on whether there are any other heating mechanisms \citep{bbf2003, dkw2004, bla2014, hwt2013}. The cooling of this gas would allow for the eventual formation of stars. 

Mergers add stellar mass to the galaxy by forming a spheroid or bulge of merged stars and creating a short-lived starburst phase in the disc \citep{t1977, b1988, v1990, spf2001, nb2003, bjc2007}. In addition, the supermassive black hole grows significantly during a merger due to the accretion of cold gas as well as the satellite's black hole merging with that of the central \citep{kh2000}. After this short-lived quasar-mode AGN feedback phase, radio-mode AGN feedback begins, where slow accretion from the hot gas atmosphere onto the supermassive black hole provides a heating source which affects the cooling of hot gas onto the disc \citep{mn2007}. We deal with AGN feedback in much greater detail in Section~\ref{sec:quenching}.

It is important to note particular differences in the prescriptions for galaxy formation physics between the H15 model and its predecessors. These changes were motivated by H15's use of Monte Carlo Markov Chain (MCMC) methods, which allowed a thorough exploration of model parameter space. They found that there was no combination of parameters in the \citet{gwb2011} model which would result in reasonable agreement with the observed stellar mass functions over the redshift range $0 \leq z \leq 3$ \citep{hwt2013}. This motivated significant modifications being made to the model in order to better match observations. 

The most significant change was with respect to the reincorporation of supernovae-ejected gas into the galaxy. In both \citet{gwb2011} and H15, gas is ejected into a reservoir which is eventually reincorporated into the hot halo and can then cool and condense onto the galaxy. For the \citet{gwb2011} model, the time it takes to be reincorporated into the hot halo is dependent on the halo mass and redshift. In the H15 model, however, the reincorporation time depends only on the halo mass and does not directly depend on redshift. These changes cause the reincorporation time to be longer for lower mass systems and shorter for higher mass systems in the H15 model than is the case in the \citet{gwb2011} model. See Fig. S2 in H15 for a visual representation of this effect. The new prescription produces behavior similar to that found in hydrodynamic studies by \citet{od2008} and \citet{odk2010}. This causes the abundance evolution of lower mass systems to be significantly different in H15 -- whereas there was very little late-time abundance evolution at low stellar masses in \citet{gwb2011}, there is a more significant change in H15 that better matches these observations. 

In order to build intuition about how to interpret observational datasets, H15 also incorporates a redshift-dependent error which models the observational errors in measuring stellar mass. This effect is included when comparing the model to the observational stellar mass functions used in their MCMC methods. The error is modeled by a Gaussian with a dispersion $0.08 (1+z)$ centred on the log of stellar mass, $\log_{10}M_{*}$. In this paper, we impose this scatter on all stellar masses used unless otherwise noted. We do this in order to visualize how we might observe galaxy growth in the real universe with the observational errors included. We note that this introduces some unphysical effects where the stellar mass of some galaxies appears to decrease from one redshift to the next. This effect is small, however, and affects a minority of galaxies in our study. 

Other changes to the model include a lower gas surface density threshold for star formation, the elimination of ram pressure stripping effects on satellites that fall into haloes with $M_{\rm{h}} < 10^{14}$ M$_{\odot}$, and an AGN feedback model that heats gas and suppresses cooling more effectively at low redshifts. For a more detailed explanation of these changes see \citet{hwt2015} and \citet{hwt2013}.

In this paper, we extensively use dark matter halo mass as an important parameter for characterizing the stellar mass growth history of central galaxies. The H15 model provides both the virial mass, $M_{\rm{vir}}$, and the maximum rotational velocity of the halo, v$_{\rm{max}}$, each of which can characterize the halo. In H15, the virial mass is defined as the mass within the virial radius which encloses a mean overdensity 200 times the critical value for the universe. In contrast, v$_{\rm{max}}$ is the maximum rotational velocity of the halo. In order to allow comparison with earlier work \citep{bwc2013, mnw2013}, we will characterize haloes using $M_{\rm{vir}}$ rather than v$_{\rm{max}}$ while noting that qualitatively our results do not change for the other choice.

We also note that for this analysis, we use only the MS and not the MS-II. The two give similar results for Milky Way-mass galaxies, but we use the former since it allows us to probe a larger number of galaxies than its smaller volume counterpart. This results in some resolution effects at low masses but does not greatly affect our results. We take note of this in the relevant sections.

In addition, the plots in this paper at times show the full central galaxy population rather than just Milky Way-mass galaxies. We note that in such cases we display a randomly selected, representative 0.2\% of the full simulation data in H15. For all plots showing Milky Way-mass galaxies, we show 2.5\% of this population where we choose the main (most massive) progenitors of these galaxies. For all statistical exercises, such as calculating medians and 68 percentile distributions, we use 100\% of the simulation data.

Finally, we note that H15 is currently the most successful semi-analytic model with regards to matching both the observed stellar mass functions and the star-forming/quiescent fractions out to $z \sim 2$. In terms of our goals for this paper, matching the stellar mass function is essential in understanding how the distribution of stellar mass throughout the universe changes as a function of redshift. H15 builds up the low mass end of the stellar mass function in a more realistic way than previous models. Since all galaxies by necessity pass through a stage where they were lower mass, agreement with the observed stellar mass functions should result in more realistic galaxy growth histories. Agreement with the observed star-forming/quiescent fraction of galaxies is also important since we focus on the cessation of star-forming galaxies and the growth of the quiescent fraction for main progenitors of Milky Way-mass galaxies. H15 therefore broadly reproduces the effect of quenching and quiescence on the evolution of the stellar mass function, even if the specific mechanism invoked is incorrect in detail. As a result, this model provides a strong foundation for our study of how quenching produces the scatter in the growth histories of central galaxies with stellar masses similar to that of the Milky Way.

\subsection{Star-formation Selection}
\label{sec:SFcut}

A main goal of this paper is to understand how differences between the pathways to star-forming and quiescent populations produce scatter in the growth histories of Milky Way-mass galaxies. We thus need a selection method to differentiate these two groups. In our analysis, we split the star-forming and quiescent populations in this model with a sSFR cut described by:

\begin{equation}
\label{eq:sSFRcut}
\rm{sSFR} = \frac{(1+\rm{z})}{2 t_{\rm{H}}},
\end{equation}
where sSFR is the specific star formation rate (SFR/$M_{*}$) in years$^{-1}$, z is the redshift, and t$_{\rm{H}}$ is the present-day Hubble time in years. Quiescent galaxies with sSFR $<$ 10$^{-12}$ yr$^{-1}$ were assigned an artificial value by H15 designed to approximately match observationally derived sSFR measurements \citep[e.g.][]{bcw2004, src2007, sck2010}. These low sSFRs signify either low-level star formation not well modeled by H15, or a contribution of older stellar populations and/or low-level AGN activity to observational sSFR estimates (see Section 5.2 in H15). Using this selection will allow us to identify and analyse the growth histories of these two groups.

\section{The Physics of Quenching in H15}
\label{sec:quenching}

Simulated galaxies in this model quench their star formation when heating energy from accretion onto a supermassive black hole offsets the radiation from cooling and infalling gas in a given halo. The general physical picture is that galaxy mergers result in the growth of the supermassive black hole by a combination of black hole mergers and the rapid feeding of cold gas into the black hole. Afterwards, the hot gas from the atmosphere around the galaxy is fed into the black hole through radio-mode accretion. The galaxy heats up its atmosphere via jets from the accretion onto the black hole.

There are many theories that attempt to explain the exact mechanism by which the atmosphere is heated, whether it be shocks due to an AGN jet injecting energy into the atmosphere \citep{frt2005, brd2009, rfg2011}, cosmic ray heating from the jet \citep{sps2008, go2008}, or effervescent heating from buoyant bubbles in the ICM \citep{b2001, bkc2002, rrn2004, vd2005, bk2002}, but the current status of these studies is inconclusive. In H15, AGN heating is extremely simplified where the AGN provides a heating rate which depends on the hot gas mass and the mass of the black hole. The heating counteracts the cooling and hot gas eventually condenses and falls onto the central galaxy, adding to its cold gas component and effectively fueling star formation.

In order to explore H15's AGN feedback model quantitatively, we will follow the formulation of heating and cooling rates detailed in H15 (See their Sections S1.4 and S1.10). We will first describe how the heating rate is calculated followed by a description of how the cooling rate is calculated in the two different regimes that the model takes into account. Finally, we will show that these rates can be approximately defined using only black hole mass and halo mass as variables. As a result, we will be able to calculate a rough boundary where heating exactly balances cooling on a $M_{\rm{BH}}$-$M_{\rm{h}}$ plot. We will show that this boundary coincides with the boundary between star-forming and quiescent galaxies, effectively building intuition for how galaxies quench in this model.

H15 accounts for AGN heating in Equation S26 of their supplementary material where,

\begin{equation}
\dot{M}_{\rm{heat}} \propto \frac{\dot{E}_{\rm{radio}}}{V_{\rm{vir}}^{2}},
\label{eqn:heating}
\end{equation}
and, following their Equations S24 and S25, 
\begin{equation}
\dot{E}_{\rm{radio}} \propto \dot{M}_{\rm{BH}} \propto M_{\rm{hot}} M_{\rm{BH}}.
\label{eqn:Eradio}
\end{equation}
Here, $\dot{M}_{\rm{heat}}$ is the heating rate from radio-mode feedback, $\dot{E}_{\rm{radio}}$ is the energy output rate due to radio-mode accretion onto the black hole, $V_{\rm{vir}}$ is the virial velocity of the dark matter halo, $M_{\rm{BH}}$ is the black hole mass, $\dot{M}_{\rm{BH}}$ is the mass accretion onto the black hole from radio-mode accretion, and $M_{\rm{hot}}$ is the hot gas mass. This formula is taken from Equation 10 in \citet{csw2006} except with the Hubble parameter divided out to provide more effective heating at later times. Inserting Equation~\ref{eqn:Eradio} into Equation~\ref{eqn:heating} gives:

\begin{equation}
\dot{M}_{\rm{heat}} \propto \frac{M_{\rm{hot}} M_{\rm{BH}}}{V_{\rm{vir}}^{2}}.
\end{equation}

In order to deal with cooling, the H15 model follows two modes by which cool gas can reach the central galaxy: the rapid infall regime and the cooling flow regime. The rapid infall regime generally describes lower mass and higher redshift haloes that experience the free fall of cool gas onto their central galaxy without a stand-off shock. At higher halo masses, cool gas flowing into the virial radius of the halo is shock heated to the virial temperature, contributing to a hot gaseous halo around the galaxy. The cooling flow regime describes the mode where the inner regions of this hot gas halo eventually cool onto the central galaxy by radiating away their energy.

In the cooling flow regime, the cooling rate is described by Equation S6 in H15,

\begin{equation}
\dot{M}_{\rm{cool}} \propto M_{\rm{hot}} \frac{r_{\rm{cool}}}{R_{\rm{vir}}},
\label{eqn:Mcool_big}
\end{equation}
where, following Equation S5,on

\begin{equation}
r_{\rm{cool}} \propto \left[\frac{M_{\rm{hot}} \Lambda}{T_{\rm{vir}} R_{\rm{vir}}} \right]^{1/2}.
\label{eqn:rcool}
\end{equation}
Plugging Equation~\ref{eqn:rcool} into Equation~\ref{eqn:Mcool_big} results in:

\begin{equation}
\dot{M}_{\rm{cool}} \propto \left( \frac{M_{\rm{hot}}}{R_{\rm{vir}}} \right)^{3/2} \left( \frac{\Lambda}{T_{\rm{vir}}} \right)^{1/2}.
\end{equation}
Here, $\dot{M}_{\rm{cool}}$ is the cooling rate of the hot gas atmosphere, $r_{\rm{cool}}$ is the cooling radius at which the cooling time equals the halo dynamical time, $R_{\rm{vir}}$ is the virial radius, $\Lambda$ is the cooling function that describes how gas cools, and T$_{\rm{vir}}$ is the virial temperature of the halo. We note here that the dynamical time of the halo, $t_{\rm{dyn}}$, depends only on $H(z)$ and therefore is constant for all haloes at a specified redshift, which is why we drop this term as a constant. This dependence on $H(z)$ has important consequences which we describe more fully in Section~\ref{sec:BHmassdependence}.

To simplify these expressions, we assume $T_{\rm{hot}} \propto T_{\rm{vir}}$ and  $M_{\rm{hot}} \propto M_{\rm{vir}}$. In addition we can use the fact that $M_{\rm{vir}} \propto V_{\rm{vir}}^{3} \propto R_{\rm{vir}}^{3}$ and $T_{\rm{vir}} \propto V_{\rm{vir}}^{2} \propto M_{\rm{vir}}^{2/3}$ in order to simplify $\dot{M}_{\rm{heat}}$ and $\dot{M}_{\rm{cool}}$ and express these quantities in terms of $M_{\rm{vir}}$ and $M_{\rm{BH}}$,

\begin{equation}
\dot{M}_{\rm{heat}} \propto M_{\rm{vir}}^{1/3} M_{\rm{BH}}
\end{equation}

\begin{equation}
\dot{M}_{\rm{cool}} \propto M_{\rm{vir}}^{2/3} \Lambda^{1/2}.
\end{equation}
Here, $\Lambda = \Lambda(T_{\rm{hot}},Z_{\rm{hot}})$ where the cooling function is defined by \citet{ds1993}. Taking the rough estimate that $\Lambda \propto T_{\rm{vir}}^{-0.7}$ in this temperature regime, we find that,

\begin{equation}
\dot{M}_{\rm{cool}} \propto M_{\rm{vir}}^{0.43}
\end{equation}

We can then find a relation between $M_{\rm{vir}}$ and $M_{\rm{BH}}$ in the case where $\dot{M}_{\rm{heat}} = \dot{M}_{\rm{cool}}$ at the boundary where heating offsets cooling for the cooling flow regime,

\begin{equation}
\frac{\dot{M}_{\rm{heat}}}{\dot{M}_{\rm{cool}}} \propto \frac{M_{\rm{vir}}^{1/3} M_{\rm{BH}}}{M_{\rm{vir}}^{0.43}}
\end{equation}

\begin{equation}
\frac{\dot{M}_{\rm{heat}}}{\dot{M}_{\rm{cool}}} \propto M_{\rm{vir}}^{-0.097} M_{\rm{BH}} \propto \rm{const}.
\end{equation}
The resulting expression is:

\begin{equation}
M_{\rm{BH}} \propto M_{\rm{vir}}^{0.097}
\label{boundary_coolingflow}
\end{equation}
for haloes in the cooling flow regime.

In the rapid infall regime, the cooling rate is described by Equation S7 in H15:

\begin{equation}
\dot{M}_{\rm{cool}} \propto M_{\rm{hot}} \propto M_{\rm{vir}}.
\end{equation}
Doing a similar exercise as with the cooling flow regime, we can find the relation between the black hole mass and the virial mass where $\dot{M}_{\rm{heat}} = \dot{M}_{\rm{cool}}$,

\begin{equation}
\frac{\dot{M}_{\rm{heat}}}{\dot{M}_{\rm{cool}}} \propto M_{\rm{vir}}^{-2/3} M_{\rm{BH}} \propto \rm{const}
\end{equation}

\begin{equation}
M_{\rm{BH}} \propto M_{\rm{vir}}^{2/3}.
\label{boundary_rapidinfall}
\end{equation}

\begin{figure}
\includegraphics[width=8.5cm]{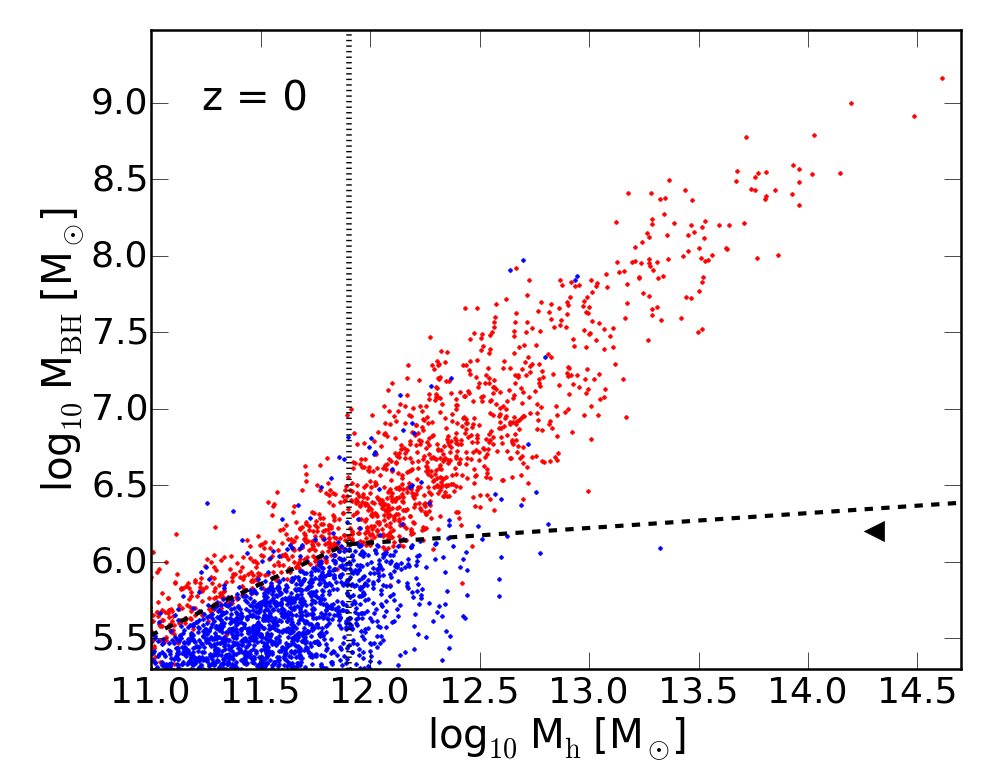}
\caption{Black hole mass as a function of halo mass for 0.2\% of all central galaxies at $z = 0$ in H15. Each point represents a galaxy where blue and red indicate star-forming and quiescent galaxies, respectively. The dashed lines represents the \textit{heating-cooling equilibrium boundary} described in Section~\ref{sec:quenching}. The vertical dotted line represents the approximate transition between two modes of gas cooling: a rapid infall and a cooling flow regime. The black arrow points to the black hole mass below which resolution effects begin to show up as differences between the black hole mass functions in the MS (shown here) and MS-II.}
\label{fig:MbhMhalo_hc}
\end{figure}

\begin{figure*}
\includegraphics[width=16.5cm]{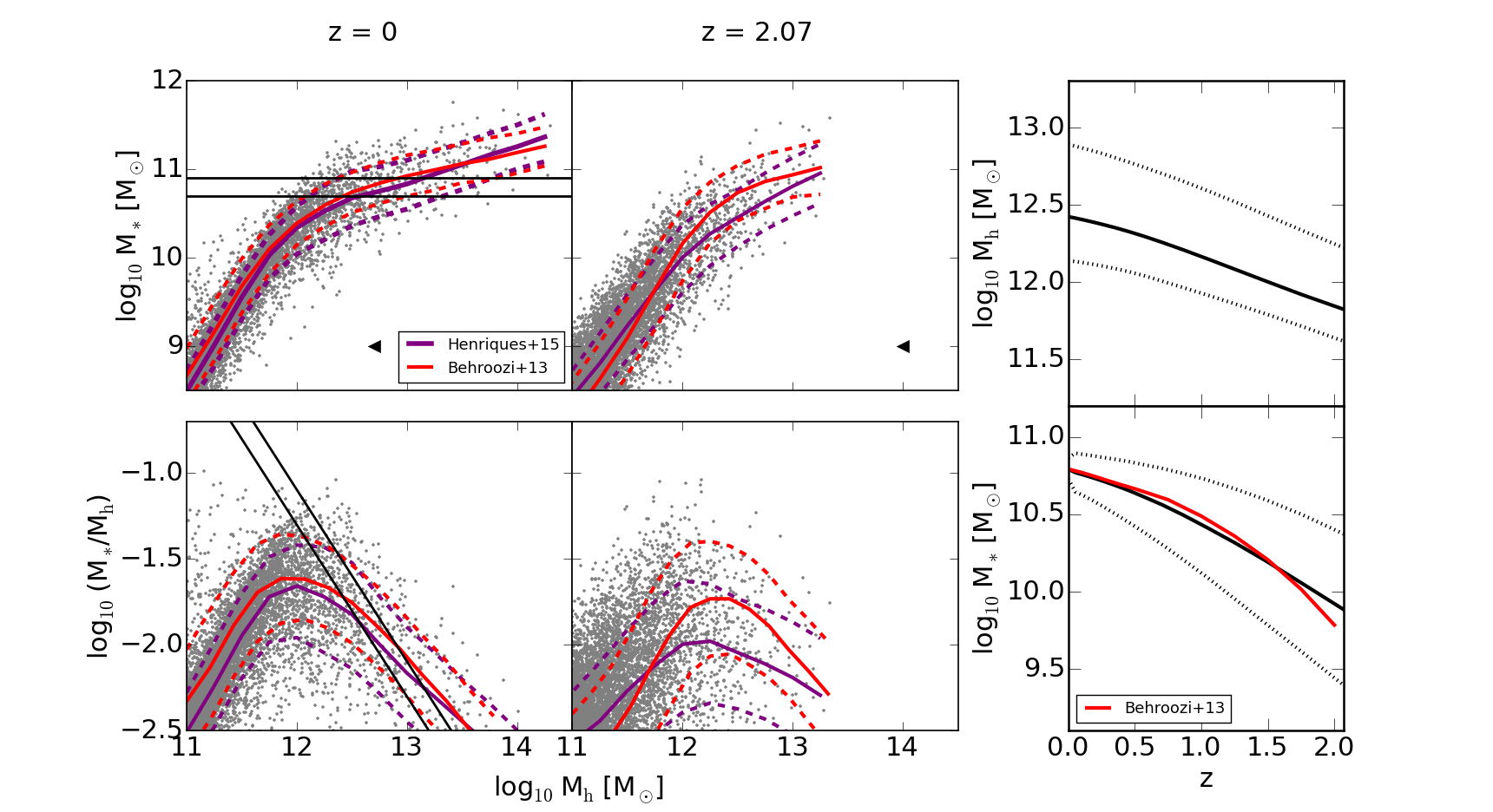}
\caption{\textit{Left}: The SMHM relation at $z = 0$ and $z = 2.07$ for 0.2\% of all central galaxies in H15. The upper two plots show stellar mass, $M_{*}$, as a function of halo mass, $M_{\rm{h}}$, and the bottom two plots show $M_{*}/M_{\rm{h}}$ as a function of $M_{\rm{h}}$. Each gray point represents one galaxy in the H15 model. The purple lines represent the median (solid) and 68 percentile scatter (dashed) of 100\% of the central galaxy population in H15, respectively. The red lines represent the median (solid) and 68 percentile scatter (dashed) of data from the \citet{bwc2013} model, respectively. The black solid lines represent the range of Milky Way-mass stellar masses we use in this paper. The black arrows point to the stellar mass below which resolution effects begin to take place in the MS. \textit{Right}: The upper and lower panels represent the halo and stellar mass tracks for all main progenitors of Milky Way-mass galaxies out to $z = 2.07$ for the H15 model, respectively. The solid black line in each plot shows the median at each redshift while the dotted black lines encompass 68 per cent of the tracks. The red line in the plot showing stellar mass tracks is the median progenitor track from the \citet{bwc2013} model.}
\label{fig:MWEvol_all}
\end{figure*}

Equations~\ref{boundary_coolingflow} and~\ref{boundary_rapidinfall} represent the slopes of the approximate boundaries between galaxies that are dominated by heating via AGN radio-mode feedback and galaxies that are dominated by cooling in one of two regimes on a $M_{\rm{BH}}$-$M_{\rm{h}}$ plot. 

In Figure~\ref{fig:MbhMhalo_hc} we show the $M_{\rm{BH}}$-$M_{\rm{h}}$ plot for a randomly-selected, representative 0.2\% of the entire central galaxy population at $z = 0$ in H15. Using the selection criteria described in Section~\ref{sec:SFcut}, we show the star-forming and quiescent populations in blue and red dots, respectively. The black arrow points to the black hole mass below which the MS's black hole mass function begins to differ from that of the MS-II due to resolution effects. The dotted vertical line represents the approximate boundary between the rapid infall regime at low halo masses and the cooling flow regime at high halo masses. Finally, we include the slopes we analytically derived above with the black dashed lines for the two different regimes. This line represents the area where AGN heating balances the cooling of hot gas onto the central galaxy, or what we will call the \textit{heating-cooling equilibrium boundary}.

We note that black hole mass and halo mass are broadly correlated. In addition, we see a fairly clear boundary between those that are star-forming and those that are quiescent. This boundary between blue and red dots is described quite well by our analytic approximation of the slopes shown with dashed lines. We note that the boundary is in the same location whether we use the MS or MS-II, regardless of the resolution effects below black hole masses $\sim$10$^{6.2}$ M$_{\odot}$. It is clear that, to a good approximation, a galaxy is quenched once heating via AGN feedback dominates over gas cooling.

This behavior is characteristic of the H15 model, but we assert that any model which simulates the cessation of star formation by balancing AGN energy input against cooling will result in a qualitatively similar behavior. In these models the cessation of star formation would be a function of black hole mass and halo mass, similar to H15's formulation.

While the populations of quiescent and star-forming galaxies are quite distinct, there is a small number of star-forming galaxies that lie above the heating-cooling equilibrium boundary. These galaxies account for about 1.3\% of the star-forming galaxy population. This scatter is a result of the time-scale in which AGN feedback effectively quenches a galaxy. These galaxies have either recently grown their black hole by a large amount or were already quiescent with a large black hole but were `revived' briefly in terms of star formation activity as a result of a merger with a gas-rich companion. The fact that so few data points overlap between the two populations in this plot demonstrates that quenching via AGN heating occurs on fairly short time-scales. It takes $\sim$0.5--1.5 Gyr for most Milky Way-mass galaxies to quench to a sSFR an order of magnitude below our sSFR cut (See Section~\ref{sec:SFcut}). Additional scatter is introduced since we made several simplifying assumptions in the above formulation of these boundaries that may fail in significant and various ways for individual galaxies. 

Even with this scatter, there is excellent agreement between our analytic approximation of the slopes for the equilibrium boundary and the actual boundary between star-forming and non-star forming galaxies. As a result, we see that halo mass and black hole mass work together on relatively short time-scales in order to prevent hot gas from cooling onto the galaxy and forming stars. We note that in H15 AGN feedback does not fundamentally depend on stellar mass. Therefore, unless there is a unique mapping between stellar mass and either halo mass or black hole mass, a galaxy's stellar mass may not provide a good characterization of its star formation properties. This has the potential to introduce a significant amount of variation in the growth histories of quenched galaxies. In the following sections, we will address specifically how the stellar mass growth is parametrized first in terms of the halo mass and finally in terms of the black hole mass. 

\section{Scatter in Milky Way-mass Galaxy Growth Histories}
\label{sec:MWgrowth}

Now that we have described the physical drivers of galaxy growth and quenching in H15, we turn to how Milky Way-mass galaxies grow. We begin by exploring the evolution of the relation between the stellar masses of these systems and their halo masses. An increasingly popular tool in this respect has been the stellar mass-halo mass (SMHM) relation which maps these two parameters onto one another \citep{ymv2009, cw2009, gwl2010, msm2010, bcw2010, ltb2012, bwc2013}. The amount of scatter in this relation quantifies the variety of stellar masses that can be contained within a halo of a given mass or the variety of halo masses which can host a central galaxy of a certain stellar mass. Halo occupation techniques that link these two parameters in order to create empirical models for the galaxy distribution have relied on the assumption that this mapping is simple. In this section, we begin to explore and challenge this assumption by analysing in detail the scatter in this relation within the H15 model. In addition, we will use Milky Way-mass galaxy growth histories in the context of this model as a case study to further explore how individual galaxies evolve with respect to the SMHM relation. 

\subsection{The Stellar Mass-Halo Mass Relation}
\label{sec:SMHMrelation}

The top two panels on the left of Figure~\ref{fig:MWEvol_all} show the stellar mass plotted against the halo mass of 0.2\% of all central galaxies in the H15 simulation for $z = 0$ and $z = 2.07$, where gray dots denote individual central galaxies in H15. The purple lines represent the median (solid) and 68 percentile scatter (dashed) of these data, calculated using 100\% of the entire central galaxy population to have accurate values for the median and the scatter. We note that this plot includes observational errors in the stellar masses as described in Section~\ref{sec:Henriques}. 

First we note the emergence of a nearly constant stellar mass population towards high halo masses at $z = 0$. In order to quantify the scatter in stellar mass at specific halo masses at $z = 0$ for H15, we provide a list of values for the SMHM relation with their associated 68 percentile scatter in Table~\ref{tab:SMHM}. We also provide the 68 percentile of the scatter for the SMHM relation without the observational scatter discussed in Section~\ref{sec:Henriques}. We find that while the scatter in stellar mass is relatively uniform for all halo masses, the scatter in halo mass increases strongly with increasing stellar mass. This signals a change in stellar mass build-up for central galaxies as they move into different mass regimes. The significant scatter in this relation is important to note when performing studies of galaxy growth since it likely originates from physical mechanisms, such as quenching, which can affect stellar mass growth within dark matter haloes.

In order to visualize the efficiency of stellar mass build-up of different sized haloes, we plot the stellar mass-halo mass ratio against the halo mass at $z = 0$ and 2.07 in the bottom two panels on the left of Figure~\ref{fig:MWEvol_all}. At low masses, the ratio of stellar mass to halo mass grows with halo mass up to about $\sim$10$^{12}$ M$_{\odot}$, at which point the ratio decreases. This turnover point has often been defined as the halo mass at which the star formation efficiency peaks. Previous studies have attempted to pinpoint a certain halo mass or stellar mass-halo mass ratio at which this peak star formation efficiency occurs in order to better understand how it evolves \citep{cw2009, ltb2012, mnw2013, bwc2013, dld2015}. If we choose $10^{12}$ M$_{\odot}$ to be the halo mass at which the relation `turns over' at $z = 0$, we see in the bottom left panel of Figure~\ref{fig:MWEvol_all} that while some of these mid-sized haloes have been relatively efficient at building up their stellar mass, many have low $M_{*}/M_{\rm{h}}$ ratios. At this halo mass, we find $\log_{10}(M_{*}/M_{\rm{h}}) = -1.66\substack{+0.24 \\ -0.30}$. Given this large scatter, defining just one point at which the efficiency peaks is a poor characterization of galaxy assembly histories. 

In order to compare the scatter in H15's SMHM relation to another model, we also show the median and 68 percentile range of the SMHM relation for the \citet{bwc2013} model (solid and dashed red lines, respectively) in both of these visualizations. \citet{bwc2013} carried out a careful, empirically-motivated analysis with the goal of estimating the SMHM relation and its scatter as a function of redshift from $z = 8$ to the present day. In addition to the intrinsic sources of scatter implemented by \citet{bwc2013}, we add observational scatter to this model's stellar masses as per Equation 11 in \citet{bwc2013}. Even though by $z = 2$ these models differ significantly, we find that the scatter in the empirical model is comparable to the scatter we find in the H15 model.

\begin{table}
\centering
\caption{The first and second columns show halo mass and stellar mass values for all central galaxies at $z = 0$ in the H15 model. The third and fourth columns show the 68 percentile scatter with and without added observational scatter, respectively.}
\label{tab:SMHM}
\begin{tabular}{llll}
\hline
\multicolumn{1}{|l|}{log$_{10}$ M$_{\rm{h}}$ [M$_{\odot}$]} & \multicolumn{1}{l|}{log$_{10}$ M$_{*}$ [M$_{\odot}$]} & \multicolumn{1}{l|}{$\sigma$ w/ O.S.} & \multicolumn{1}{l|}{$\sigma$ w/o O.S.} \\ \hline
11                                                       & 8.49                                                   & $\substack{+0.23 \\ -0.20}$                          & $\substack{+0.22 \\ -0.18}$                           \\
11.25                                                  & 9.01                                                   & $\substack{+0.25 \\ -0.24}$                          & $\substack{+0.25 \\ -0.22}$                           \\
11.5                                                    & 9.55                                                   & $\substack{+0.24 \\ -0.25}$                          & $\substack{+0.23 \\ -0.23}$                           \\
11.75                                                  & 10.03                                                 & $\substack{+0.23 \\ -0.26}$                          & $\substack{+0.22 \\ -0.25}$                           \\
12                                                       & 10.34                                                 & $\substack{+0.24 \\ -0.30}$                          & $\substack{+0.22 \\ -0.29}$                           \\
12.25                                                  & 10.53                                                 & $\substack{+0.29 \\ -0.33}$                          & $\substack{+0.28 \\ -0.32}$                           \\
12.5                                                    & 10.68                                                 & $\substack{+0.30 \\ -0.31}$                          & $\substack{+0.29 \\ -0.30}$                           \\
12.75                                                  & 10.75                                                 & $\substack{+0.28 \\ -0.29}$                          & $\substack{+0.26 \\ -0.27}$                           \\
13                                                       & 10.84                                                 & $\substack{+0.27 \\ -0.28}$                          & $\substack{+0.25 \\ -0.27}$                           \\
13.25                                                  & 10.95                                                 & $\substack{+0.26 \\ -0.28}$                          & $\substack{+0.24 \\ -0.27}$                           \\
13.5                                                    & 11.06                                                 & $\substack{+0.24 \\ -0.28}$                          & $\substack{+0.23 \\ -0.27}$                           \\
13.75                                                  & 11.16                                                 & $\substack{+0.24 \\ -0.27}$                          & $\substack{+0.22 \\ -0.26}$                           \\
14                                                       & 11.26                                                 & $\substack{+0.25 \\ -0.25}$                          & $\substack{+0.23 \\ -0.24}$                           \\
14.25                                                  & 11.37                                                 & $\substack{+0.26 \\ -0.28}$                          & $\substack{+0.24 \\ -0.25}$                          
\end{tabular}
\end{table}

\begin{figure*}
\includegraphics[width=16cm]{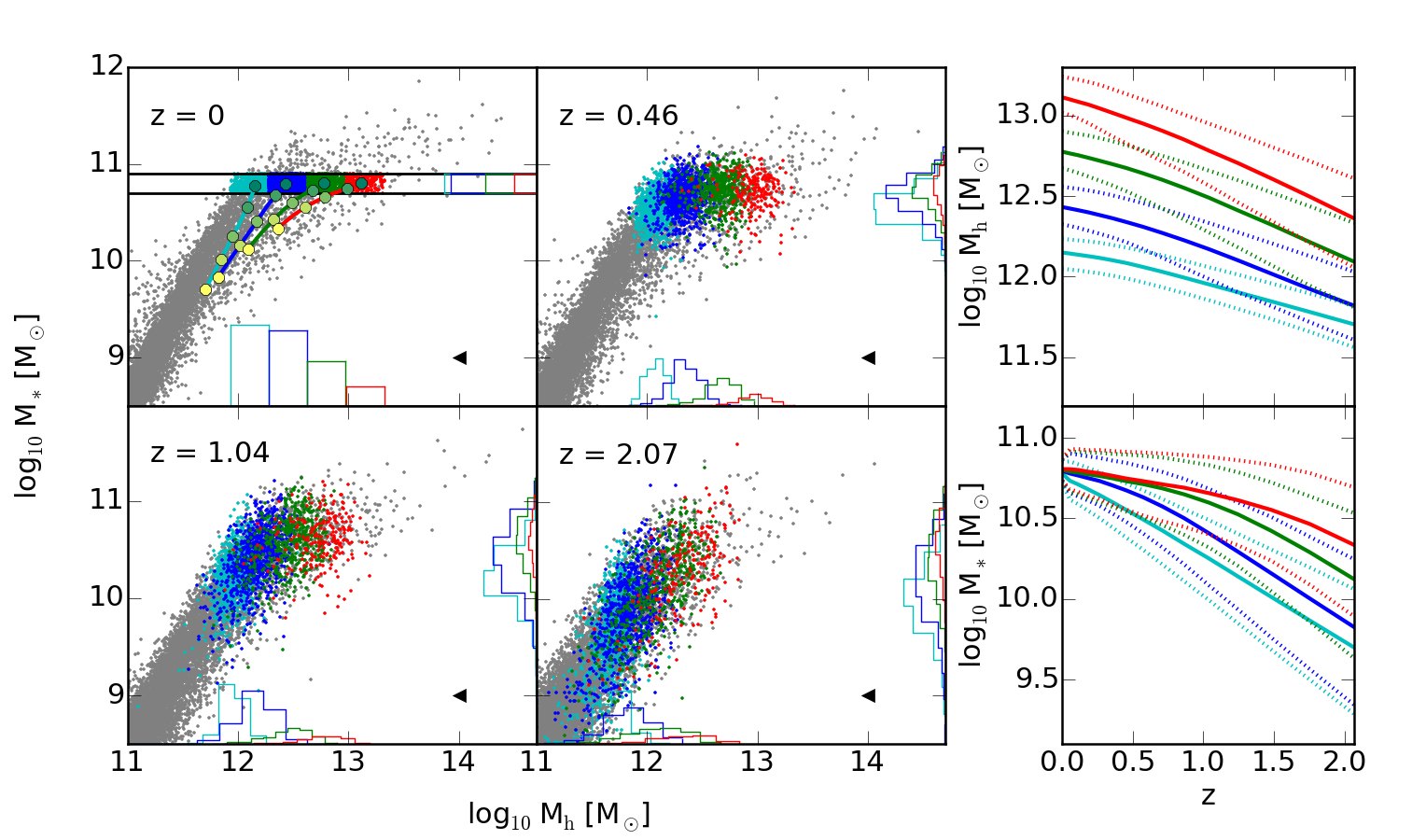}
\caption{\textit{Left}: The SMHM relation for Milky Way-mass galaxies at $z = 0$, 0.46, 1.04, and 2.07, where the gray dots represent the 0.2\% of the total galaxy population in H15. The colored points represent 2.5\% of all main progenitors of present-day Milky Way-mass galaxies split into four halo mass bins, where the boundaries are at $M_{\rm{h}}$ = 10$^{12.5}$, 10$^{12.9}$, and 10$^{13.3}$ M$_{\odot}$. Each panel also shows the histogram of the distributions for each group in stellar mass and halo mass on the right and bottom edges of the plots, respectively. These histograms are scaled with respect to the total number of Milky Way-mass galaxies. In the $z = 0$ panel, we track the evolution of the median stellar and halo mass of the main progenitors of these galaxies in time. The large colored circles represent the median values at $z = 0$, 0.46, 1.04, 1.48, and 2.07 where the circles go from green to yellow with increasing redshift. The black arrows point to the stellar mass below which resolution effects begin to take place in the MS. \textit{Right}: The upper and lower panels represent the halo and stellar mass tracks for all main progenitors of Milky Way-mass galaxies split into four halo mass bins at $z = 0$. The solid line in each plot shows the median at each redshift while the dotted lines encompass 68 per cent of the tracks.}
\label{fig:Evoltracks}
\end{figure*}

\subsection{The General Population of Milky Way-mass Galaxies}
\label{sec:GeneralPopulation}

We now turn to Milky Way-mass galaxies and their main progenitors as a case study to better understand the evolution of the SMHM relation, especially in the context of its scatter. This exercise will also help gain insight into the variety of pathways that result in a central galaxy with the stellar mass of the Milky Way at $z = 0$. 

The solid black lines in the leftmost panels of Figure~\ref{fig:MWEvol_all} indicate lines of constant stellar mass at 5\e{10} and 8\e{10} M$_{\odot}$. In the rightmost panels of Figure~\ref{fig:MWEvol_all}, we take all central galaxies at $z = 0$ with stellar masses in this range and trace out the median and 68 percentile range of halo and stellar masses of their main progenitors out to $z = 2.07$ in the upper and lower panels, respectively. We note that these plots include observational errors in stellar mass as described in Section~\ref{sec:Henriques}.

\begin{figure*}
\includegraphics[width=16cm]{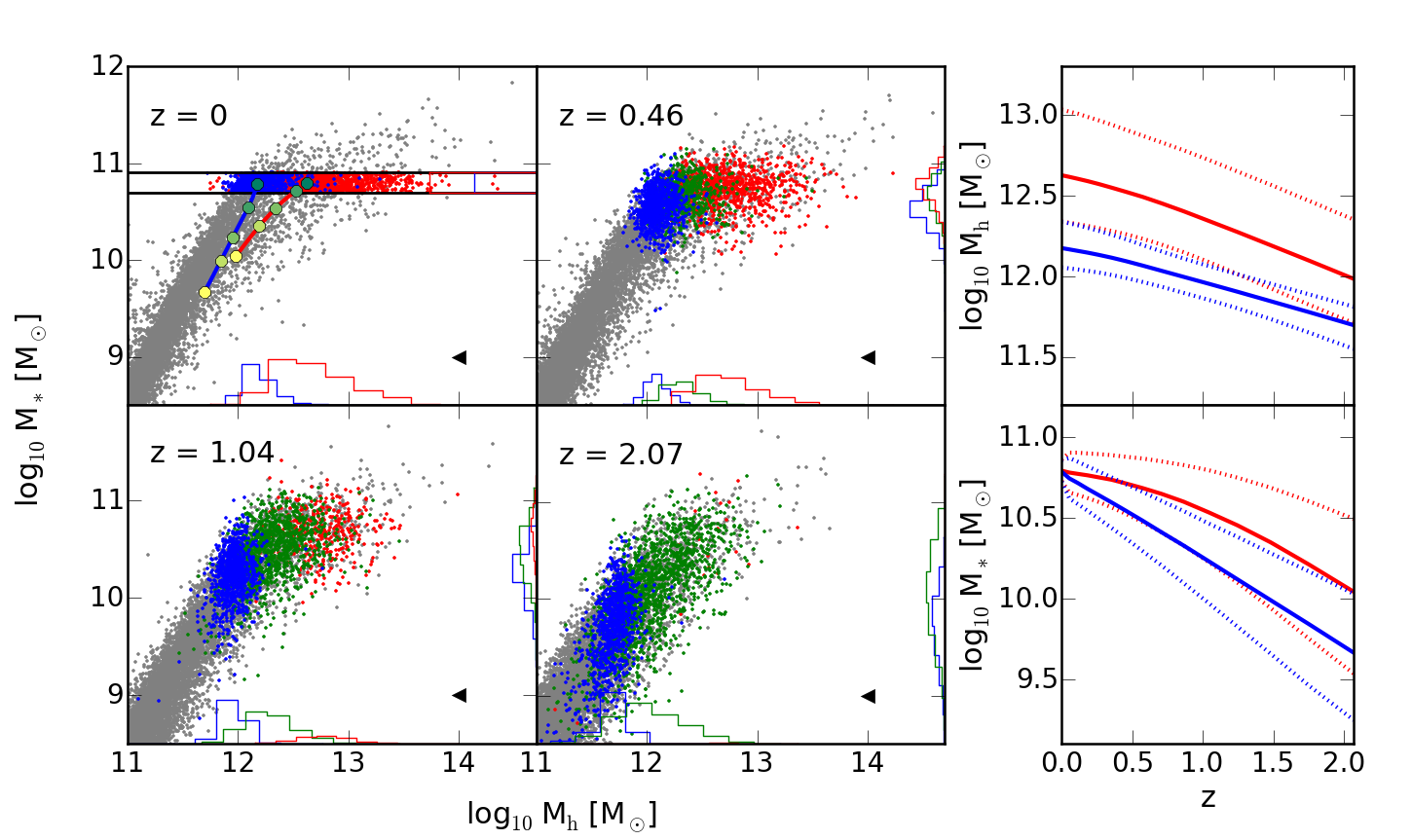}
\caption{\textit{Left}: The SMHM relation for Milky Way-mass galaxies at $z = 0$, 0.46, 1.04, and 2.07, where the gray dots represent the 0.2\% of the total galaxy population in H15. The colored points represent 2.5\% of all main progenitors of present-day Milky Way-mass galaxies, where blue dots represent galaxies that have remained star-forming since $z = 2.07$, green dots represent galaxies that are star-forming at the given redshift but will become quiescent by the present day, and red dots represent galaxies that have quenched and that will remain quiescent up to the present day. Each panel also shows the histogram of the distributions for each group in stellar mass and halo mass on the right and bottom edges of the plots, respectively. These histograms are scaled with respect to the total number of Milky Way-mass galaxies. In the $z = 0$ panel, we track the evolution of the median stellar and halo mass of the main progenitors of galaxies that have always been star-forming in blue and galaxies that become quiescent by the present day in red. The large colored circles represent the median values at $z = 0$, 0.46, 1.04, 1.48, and 2.07 where the circles go from green to yellow with increasing redshift. The black arrows point to the stellar mass below which resolution effects begin to take place in the MS. \textit{Right}: The upper and lower panels represent the halo and stellar mass tracks for all main progenitors of present-day star-forming and quiescent Milky Way-mass galaxies in blue and red, respectively. The solid line in each plot shows the median at each redshift while the dotted lines encompass 68 per cent of the tracks.}
\label{fig:Evoltracks_rb}
\end{figure*}

First, we note a large range of halo masses for Milky Way-mass galaxies at all redshifts, a trend also seen in the upper leftmost panel where the scatter in the SMHM relation at Milky Way masses is quite large. While the 68 percentile range in halo masses stays roughly the same with a modest 0.15 dex decrease at high redshifts, our narrow range of present-day stellar masses increases from 0.13 to 0.98 dex by $z = 2.07$. This shows that there is a wide diversity of ways to become a Milky way-mass galaxy, where some grow more than twenty times their stellar mass while others grow very little.

One of our goals is to understand the origin of this diversity of growth histories. In Figure~\ref{fig:Evoltracks}, we show the SMHM relation at $z = 0$, 0.46, 1.04, and 2.07. The gray dots represent 0.2\% of all central galaxies and the colored dots show 2.5\% of all main progenitors of Milky Way-mass galaxies at the present day. The different colors within this latter group represent Milky Way-mass galaxies within different halo mass ranges at $z = 0$. Evolving them backwards into their main progenitor stellar and halo masses results in the distributions shown both in the colored dots and in the histograms in each panel. The median evolutionary tracks of each halo mass bin are shown in the top left panel in colored lines where each circle represents the median value at $z = 2.07$, 1.48, 1.04, 0.46, and 0, color coded from yellow to green, respectively. 

We note that galaxies with different present-day halo masses exhibit different median growth tracks on the SMHM relation. Galaxies with less massive haloes tend to grow in stellar mass rapidly towards the present day, whereas galaxies with more massive haloes had already formed a large fraction of their present-day stellar mass by $z = 2.07$ and grew very little afterwards. This is more clearly shown in the rightmost panels of Figure~\ref{fig:Evoltracks}, where we plot the halo and stellar mass evolution of these four Milky Way-mass galaxy groups split by halo mass in the upper and lower panels, respectively. As before, the solid lines indicate medians and the dotted lines indicate the 68 percentile scatter. This illustrates an important finding of our study -- there is a great deal of correlation between growth history and halo mass in the H15 model, in the sense that more massive haloes tend to grow most of their stellar mass earlier. Such `anti-hierarchical' behavior has long been inferred from observational datasets \citep[e.g.][]{tmb2005} and emerges naturally from the H15 model.

While stellar mass at high redshift does correlate slightly with growth history for Milky Way-mass galaxies, as is shown in the vertical histograms, we also see that halo mass does a much better job at differentiating between distinct pathways of galaxy growth in the SMHM relation. These trends will provide the basis upon which we will continue our study of this galaxy population with respect to their star formation activity.

\section{The Star-forming and Quiescent Populations of Milky Way-mass Galaxies}
\label{sec:SFQPopulations}

Now that we have quantified the scatter in the stellar and halo mass growth histories of Milky Way-mass galaxies, we use the selection criteria described in Section~\ref{sec:SFcut} to divide our Milky Way-mass galaxy sample into star-forming and quiescent galaxies at the present day. Complementary to the right panel of Figure~\ref{fig:MWEvol_all}, the right panel of Figure~\ref{fig:Evoltracks_rb} shows the median and 68 percentile range of halo and stellar masses of the main progenitors of these groups out to $z = 2.07$ in the upper and lower panels, respectively. In our sample, 36\% of galaxies are star-forming at $z = 0$ while 64\% are quiescent. Although these two populations do exhibit significant differences in growth histories, we note the large overlap in stellar masses between the present-day star-forming and quiescent populations. In contrast, the growth histories of halo masses are quite distinct, where haloes that host galaxies that will quench by $z = 0$ are generally more massive at all redshifts.

A main goal of our study is to understand the scatter of stellar mass growth histories of Milky Way-mass galaxies. While splitting our sample population into star-forming and quiescent descendants does hint at the origin of much of this scatter, there is a significant amount of it still unaccounted for in the evolutionary tracks of these two groups. We first comment on the scatter in star-forming galaxies' growth histories and then focus on that of the quiescent galaxies' growth histories.

\subsection{Scatter in Growth Histories of Star-Forming Galaxies}
\label{sec:scatterSFgalaxies}

Figure~\ref{fig:Evoltracks_rb} shows significant scatter in stellar mass growth histories of currently star-forming Milky Way-mass galaxies, albeit less so than those which are currently quiescent. A good deal of this scatter originates from `observational error' in $M_{*}$ that we impose on the `true' stellar masses to match the observed stellar mass functions, as discussed in Section~\ref{sec:Henriques}. Removing these `observational stellar mass errors' diminishes the total range of these progenitor stellar masses from 0.78 to 0.58 dex at $z = 2.07$ and 0.49 to 0.32 dex at $z = 1.04$. While the scatter in growth histories from `observational error' does not reflect true changes in the mass of the model galaxies, it is crucial to account for in studies attempting to connect galaxy populations at different cosmic epochs.

The remaining amount of scatter in the tracks for star-forming galaxies shown in Figure~\ref{fig:Evoltracks_rb} comes from the physical prescriptions used in H15. The models for gas cooling, star formation, and feedback depend on the cold gas mass, the radius of the gas disc, and the dynamical time of the disc. The cold gas mass is largely a function of halo mass and halo growth history, the radius of the gas disc depends on the spin parameter, and the dynamical time depends on the maximum halo velocity and therefore the concentration of the halo. Consequently, an intrinsic diversity of values for these halo quantities imposes scatter in star formation histories at a given halo mass, driving the majority of the scatter in growth histories of star-forming galaxies seen in Fig~\ref{fig:Evoltracks_rb}. We note that this physical source of scatter affects both the star-forming and quiescent populations in the model.

\begin{figure}
\includegraphics[width=8.5cm]{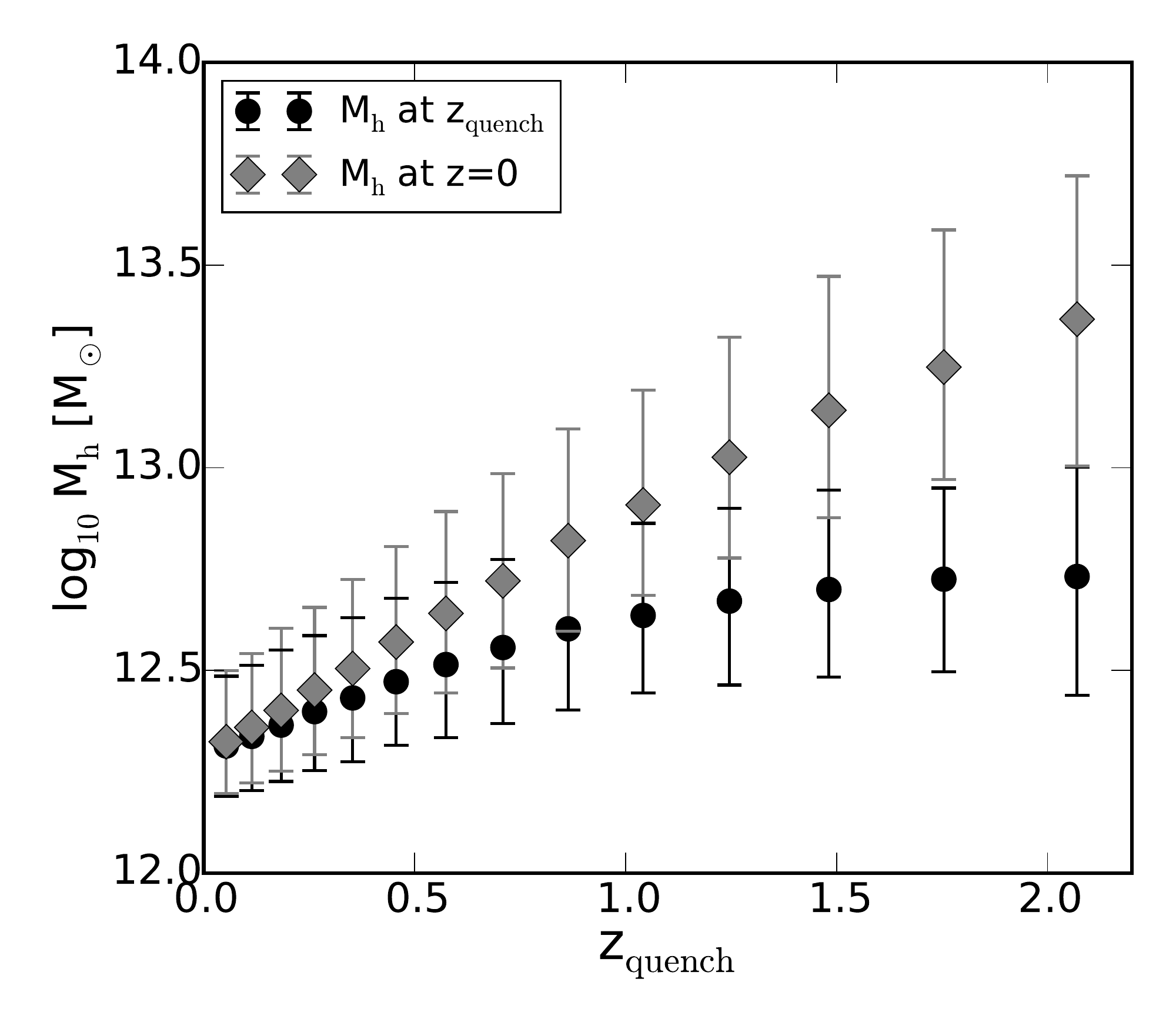}
\caption{Halo mass as a function of the quenching redshift of all Milky Way-mass galaxies in the H15 model. Grey diamonds represent the median present-day halo mass, $M_{\rm{h}}$($z = 0$), of galaxies that quenched at each redshift while black circles represent the median halo mass of those same galaxies at the redshift they quenched, $M_{\rm{h}}$($z = z_{\rm{quench}}$). The error bars represent the 68 percentile scatter in halo masses for galaxies that quenched at each respective redshift.}
\label{fig:staggeredquenching}
\end{figure}

\begin{figure}
\includegraphics[width=8cm]{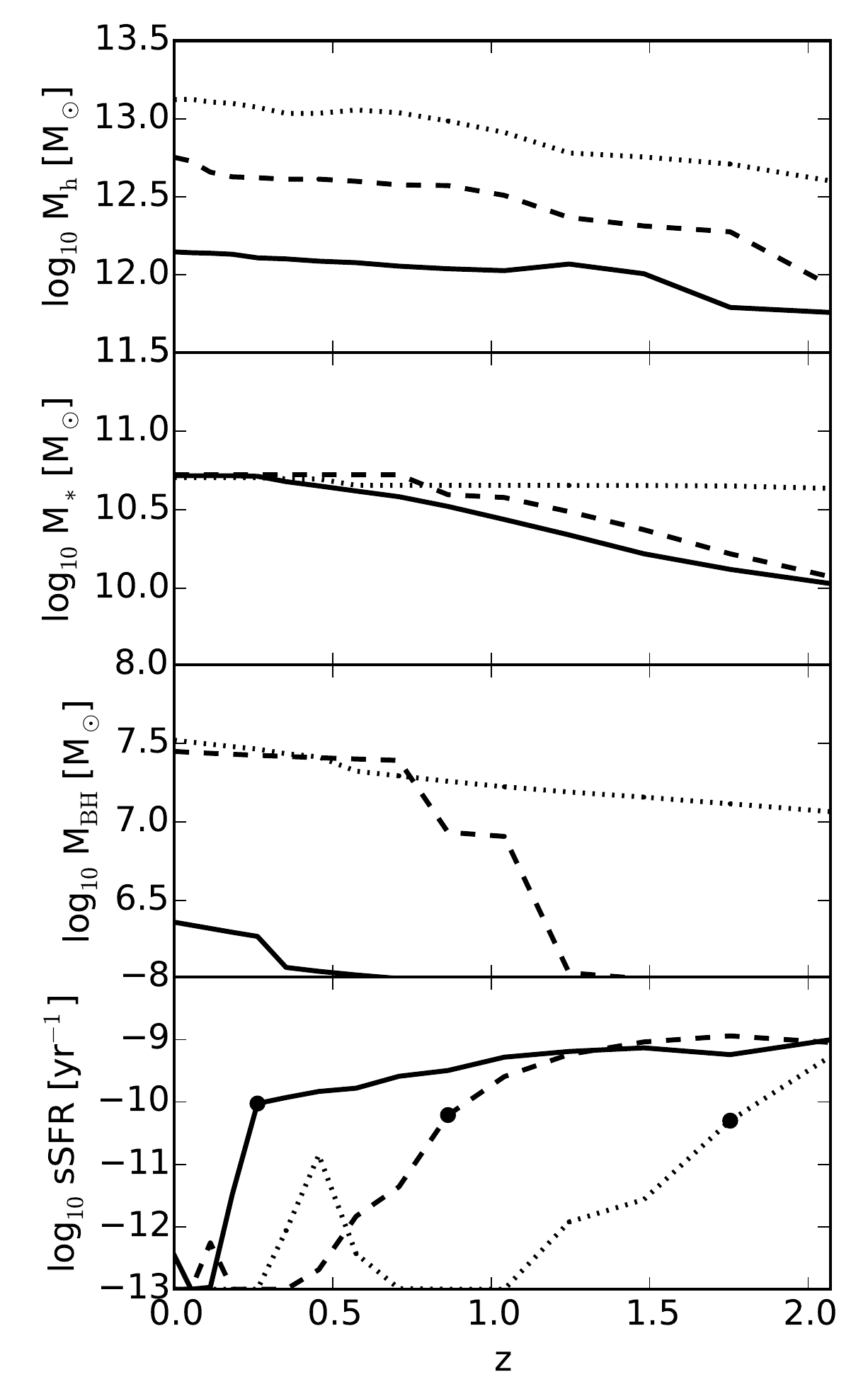}
\caption{Halo mass, stellar mass, black hole mass, and specific star formation rate tracks for three representative Milky Way-mass galaxies in the H15 model. Each galaxy is denoted by either a solid, dashed, or dotted line. The black dots in the sSFR tracks represent the last redshift at which they are star-forming according to our definition (See Section~\ref{sec:SFcut}).}
\label{fig:tracks}
\end{figure}

\subsection{Scatter in Growth Histories of Quiescent Galaxies: Staggered Quenching}
\label{sec:scatterQgalaxies}

In contrast to galaxies that have always been star-forming, the scatter in evolutionary tracks of quiescent galaxies is particularly pronounced. At one extreme, a currently quiescent galaxy could follow the evolutionary track of a star-forming galaxy up until relatively recent times, only to diverge from that track at $z < 0.5$. At the other extreme, a quiescent galaxy could follow the evolutionary track of a galaxy that grows rapidly at $z > 2.07$ and very little thereafter. This results in a large overlap region where star-forming galaxies and quiescent galaxies at the present day could have had similar stellar masses at earlier times. While star-forming galaxies do exhibit quite a bit of scatter, their growth histories match in the sense that they seem to be more steadily growing their stellar mass towards the present day, rather than exhibiting a stunted growth. In order to understand these trends we must understand how the stellar mass growth of the quiescent population is affected by the cessation of star formation.

To illustrate, we show the SMHM relation for 2.5\% of all main progenitors of present-day Milky Way-mass galaxies at $z = 0$, 0.46, 1.04, and 2.07 in the left panels of Figure~\ref{fig:Evoltracks_rb}. Blue dots represent galaxies that have remained star-forming since $z = 2.07$, green dots represent galaxies that are star-forming at the given redshift but will become quiescent by $z = 0$, and red dots represent galaxies that have quenched and that will remain quiescent up to the present day. In the top left panel of Figure~\ref{fig:Evoltracks_rb}, we plot median tracks on the SMHM relation of Milky Way-mass galaxies that remain star-forming in blue and Milky Way-mass galaxies that have become quiescent by $z = 0$ in red, where each circle represents the median value at $z = 2.07$, 1.48, 1.04, 0.46, and 0 color coded from yellow to green.

Almost all Milky Way-mass galaxies were star-forming at $z = 2.07$, even those that will become quiescent by the present day. We also note that the quiescent population grows gradually, where galaxies in more massive haloes quench earlier than those in lower mass haloes. 

In order to probe this behavior more directly, we evaluate the quenching redshift of all central Milky Way-mass galaxies that are quiescent at $z = 0$. In Figure~\ref{fig:staggeredquenching}, we plot the median and 68 percentile range of present-day halo masses against the redshift at which these galaxies quenched as gray diamonds. In black circles we plot the halo masses at the redshift at which the galaxy quenches against the quenching redshift. The halo mass difference between the gray diamonds and the black circles show us the median halo mass growth since the time of quenching. We see a scattered but clear correlation between the present-day halo masses of galaxies and the redshift at which these galaxies quench. We call this behavior \textit{staggered quenching}. This demonstrates that present-day high mass haloes tend to have quenched earlier than present-day lower mass haloes, showing that galaxies have undergone different growth histories as a function of their halo masses. 

This correlates well with the behavior we saw in Section~\ref{sec:GeneralPopulation}, where a population with a large range in halo mass points to a large diversity of growth histories. In the upper right panel of Figure~\ref{fig:Evoltracks_rb}, we note the much larger range of halo masses for those galaxies that have become quiescent by the present day and the much narrower range of halo masses for those galaxies that have remained star-forming, where once again these ranges encompass 68 per cent of the data. The larger diversity of halo masses for today's quiescent galaxies implies they also have a larger variation of growth histories than their star-forming counterparts, at least in the context of this model.

While halo mass broadly correlates with quenching times in this model, we have already shown in Section~\ref{sec:quenching} that black hole mass also plays a vital role in heating the atmospheres of galaxies via AGN radio-mode feedback. In order to explore this further, in Figure~\ref{fig:tracks} we plot halo mass, stellar mass (without observational scatter), black hole mass, and specific star formation rate tracks for three representative Milky Way-mass galaxies that quenched at $z = 0.18$, 0.7, and 1.47 in solid, dashed, and dotted lines, respectively. The black dots in the sSFR tracks indicate the last redshift at which this galaxy is classified as star-forming.

In this visualization, we see the effect of staggered quenching where galaxies in larger haloes quench earlier. While this trend exists, the significant mass growth of the central black hole for two of these systems directly coincides with a significant decrease in the galaxies' sSFR. \footnote{We note that while the galaxies that quench at $z = 0.18$ and 0.7 grow their black hole mass by a large amount in a very short time, this is not representative of all Milky Way-mass galaxies since some do become quiescent as their central black hole grows more gradually. This is the case for the galaxy that quenches at $z = 1.47$ -- its black hole is already very large and, as a result, halts its star formation early on without the need for significant black hole mass growth. We discuss this phenomenon in more detail in Section~\ref{sec:BHmassdependence}.} This demonstrates the limits of looking only at halo mass and staggered quenching in order to explain the onset of quiescence. In the following section, we focus on how black hole mass, halo mass, and quenching are connected with regards to Milky Way-mass galaxies in H15.

\begin{figure}
\includegraphics[width=8.5cm]{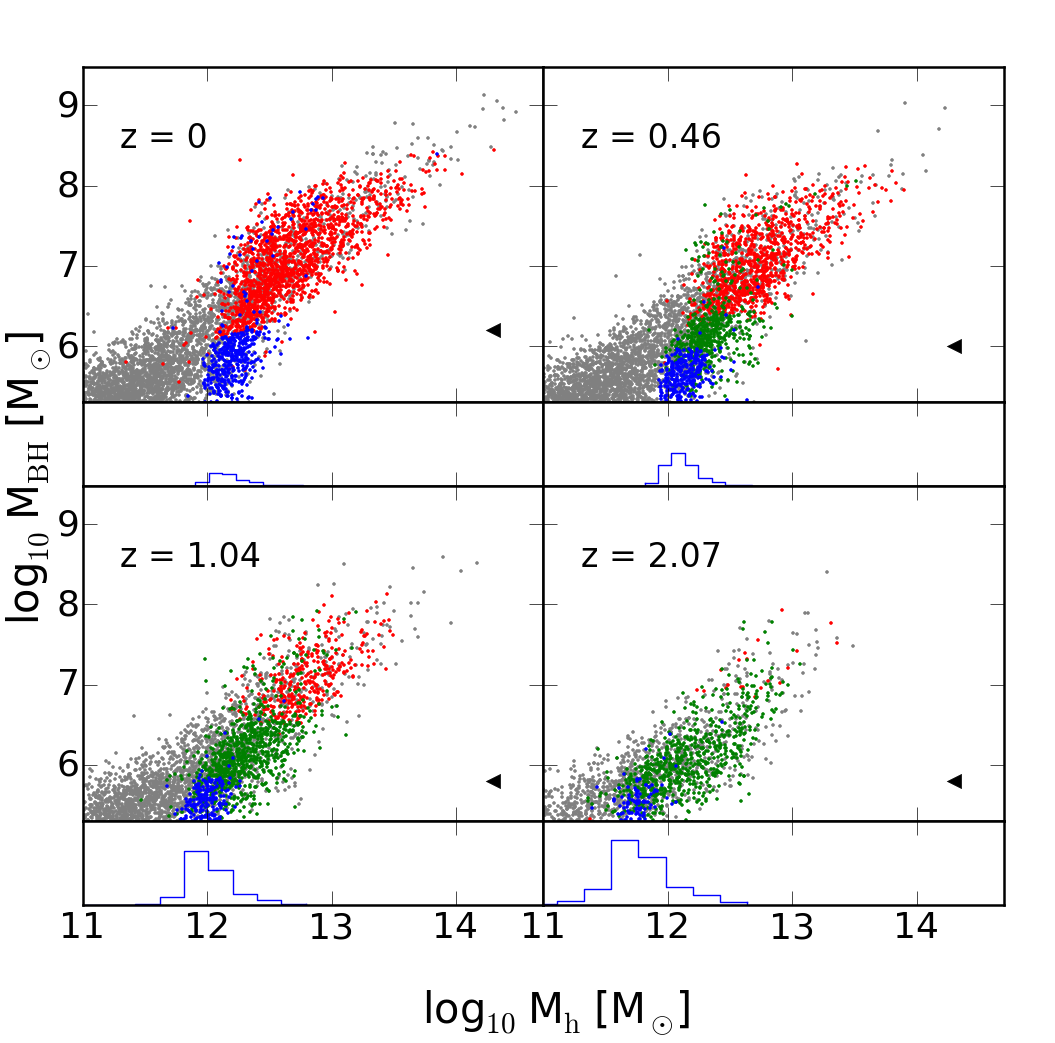}
\caption{The black hole mass-halo mass relation at $z = 0$, 0.46, 1.04, and 2.07 where gray dots represent 0.2\% of the whole central galaxy population in H15. The colored dots represent 2.5\% of all main progenitors of Milky Way-mass galaxies; blue dots represent those that have remained star-forming since $z = 2.07$, green dots represent galaxies that are star-forming at the given redshift but will become quiescent by the present day, and red dots represent galaxies that have quenched and that will remain quiescent up to the present day. The histograms represent the distribution of halo masses for those galaxies that have a black hole mass of zero. We note that these galaxies are only star-forming; quiescent central galaxies do not have zero-mass black holes.}
\label{fig:MbhMhalo_MW}
\end{figure}

\section{Black Hole Mass Dependence}
\label{sec:BHmassdependence}

In H15's framework, mergers cause most of the black hole mass growth. A supermassive black hole is first formed at the centre of a galaxy via quasar-mode feedback after a merger, whether it be major or minor. The more equal the merger ratio and the more cold gas there is in the colliding galaxies, the more massive the initial black hole. In contrast, radio-mode AGN feedback adds a negligible amount of mass onto the black hole (See \citealt{csw2006}, Figure 3). A consequence of this is that black hole growth is completely determined by the merger histories of galaxies, which are largely stochastic by nature. This connection between the black hole and halo mass of a galaxy has important implications for how Milky Way-mass galaxies quench in H15.

In order to incorporate the importance of the central black hole within our discussion, we return to the black hole mass-halo mass relation. In Figure~\ref{fig:MbhMhalo_MW}, we show black hole mass as a function of halo mass for 0.2\% of all centrals (gray) and for 2.5\% of all main progenitors of Milky Way-mass galaxies (colored) at $z = 0$, 0.46, 1.04, and 2.07. As before, blue dots represent galaxies that have remained star-forming since $z = 2.07$, green dots represent galaxies that are star-forming at the given redshift but will become quiescent by the present day, and red dots represent galaxies that have quenched and that will remain quiescent up to the present day. The black arrows point to the black hole mass below which the MS's black hole mass function begins to differ from that of the MS-II due to resolution effects. We note that even with this limitation, the results for Milky Way-mass galaxies in both the MS and MS-II are qualitatively similar. The histograms at each redshift represent the distribution of halo masses for those galaxies that have a black hole mass of zero in the MS\footnote{In the MS-II, such galaxies have low, non-zero black hole masses as a consequence of being able to resolve small mergers. This is not the case in the MS.}. We find that all central galaxies with no central black hole are star-forming, providing further evidence that the black hole is a crucial ingredient to quiescence. In addition, we note that galaxies that are star-forming but will become quiescent (green dots) occupy a region between those that will remain star-forming (blue) and those that have already become quiescent (red).

In this visualization, the heating-cooling equilibrium boundary introduced in Section~\ref{sec:quenching} can be seen as the boundary between red and blue dots at $z = 0$ and red and green dots at higher redshifts. We note that Milky Way-mass galaxies in the H15 model quench almost exclusively in the cooling flow regime, leading to a shallow boundary between heating-dominated and cooling-dominated Milky Way-mass systems on a $M_{\rm{BH}}$-$M_{\rm{h}}$ plot. This means that, at a given redshift, quiescence is primarily a function of black hole mass. The boundary evolves in time, eventually decreasing the black hole mass thresholds for quiescence at lower redshifts. The decreasing threshold is mainly because H15 designed the model so that $\dot{M}_{\rm{cool}}$ is inversely proportional to the dynamical time and the virial radius of the halo, both of which depend on the Hubble parameter, $H(z)$. This dependence on $H(z)$ results in $\dot{M}_{\rm{cool}}$ being much larger at higher redshift while the AGN heating term stays almost constant due to its weak dependence on the virial mass. While the slope of the boundary stays the same -- defined by galaxies being in the cooling flow regime at these halo masses -- the normalization changes with redshift. 

This has important implications for the central galaxy population at $z = 0$ since cooling becomes increasingly less effective at late times for these model galaxies. We note that the shallow equilibrium boundary derived in Equation~\ref{boundary_coolingflow} in Section~\ref{sec:quenching} means halo mass growth is not the fundamental reason for quenching in this model. The reason why there is a correlation between halo mass and quiescence (see Section~\ref{sec:scatterQgalaxies}) is because of the positive correlation between halo and black hole mass. As the equilibrium boundary decreases, lower mass haloes are able to quench as a result of this positive correlation. This effectively results in the staggered quenching behavior we saw in Section~\ref{sec:scatterQgalaxies}.

\begin{figure}
\includegraphics[width=8.5cm]{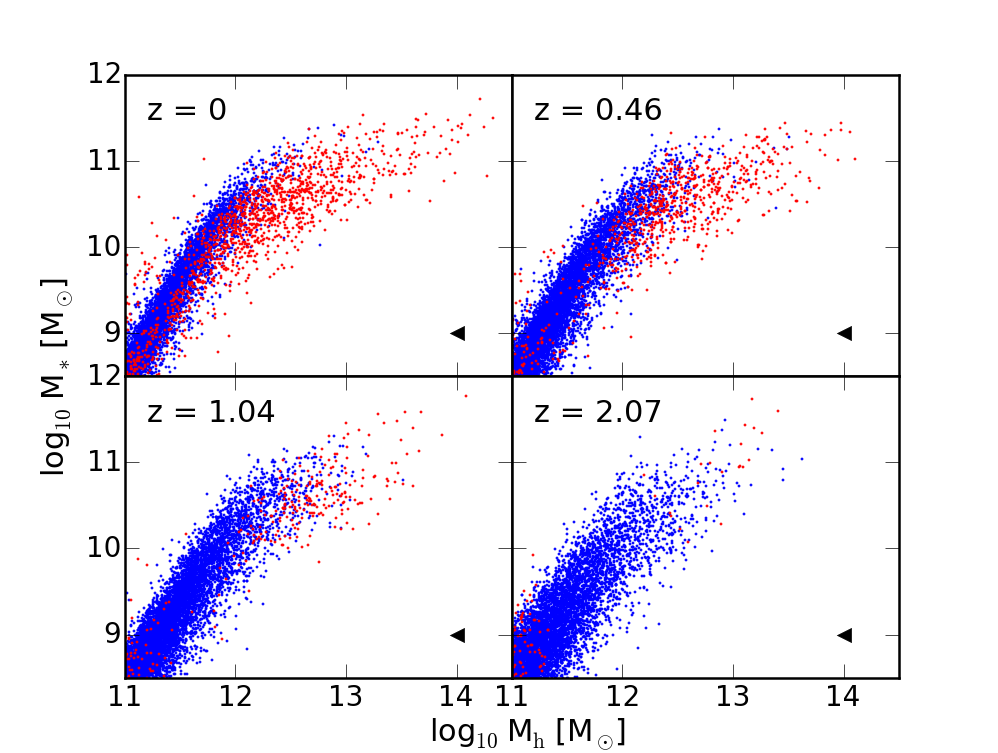}
\caption{The SMHM relation for 0.2\% of all central galaxies split into star-forming (blue) or quiescent (red) at $z = 0$, 0.46, 1.04, and 2.07. The black arrows point to the stellar mass below which resolution effects begin to take place in the MS.}
\label{fig:redblue_Evoldist}
\end{figure}

In addition, this means that a galaxy does not necessarily need to drastically increase its black hole mass in order to quench since the dependence on the Hubble parameter naturally evolves the black hole mass threshold for quiescence to lower values with decreasing redshift. This results in some Milky Way-mass systems in H15 that already have black holes large enough that they do not need to grow anymore in order to quench -- this is the case for the galaxy that quenches at $z = 1.47$ in Figure~\ref{fig:tracks}. Instead of the rapid black hole mass growth seen in the other two galaxies as a result of a major merger, this galaxy quenches mainly as a result of a decreasing cooling efficiency. At earlier times cooling is intense enough to offset AGN heating. As the efficiency of cooling drops, however, the heating effectively stops star formation. These galaxies may help interpret quiescent galaxies with large bulges but large or dominant disks that do not show evidence of a recent major merger. In terms of Milky Way-mass galaxies, we find that 65.9\% of this population shows the onset of quiescence concurrently with a black hole growth of twofold or greater and 33.2\% shows the onset of quiescence with an order of magnitude or greater growth in black hole mass. Nevertheless, for Milky Way-mass galaxies, systems which show a more gradual black hole growth into quiescence are not negligible in this model.

\begin{figure*}
\includegraphics[width=16.5cm]{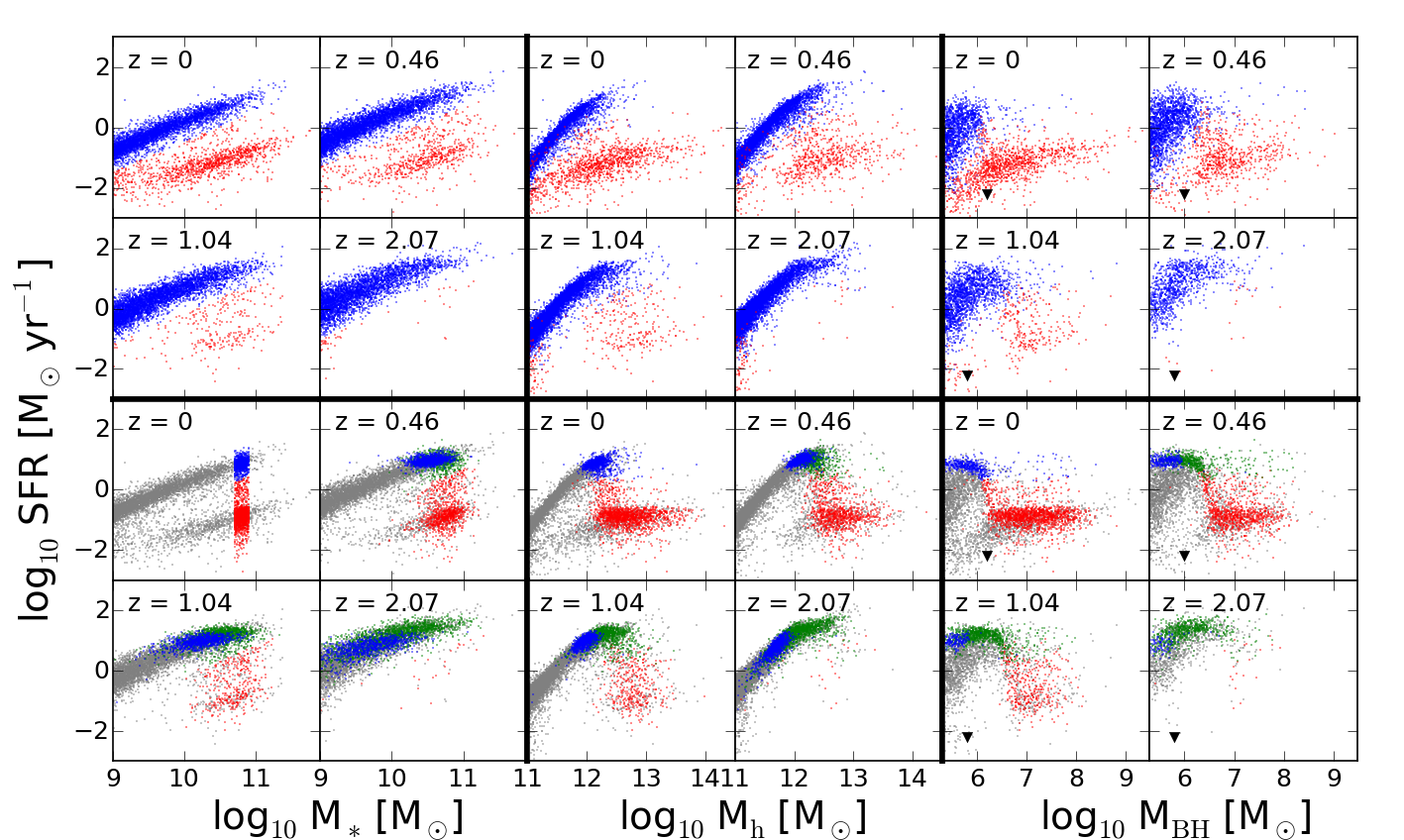}
\caption{The star formation rate (SFR) as a function of stellar mass (left), halo mass (middle), and black hole mass (right) at $z = 0$, 0.46, 1.04, and 2.07. The top panels show 0.2\% of the whole central population split into star-forming and quiescent galaxies at each respective redshift. The bottom panels show 0.2\% of the central galaxy population in gray dots as reference and 2.5\% of all main progenitors of Milky Way-mass galaxies in colored dots. Blue dots represent those Milky Way-mass galaxies that have remained star-forming since $z = 2.07$, green dots represent those that are star-forming at the given redshift but will become quiescent by the present day, and red dots represent those that have quenched and that will remain quiescent up to the present day. The black arrows point to the black hole mass below which the MS's black hole mass function begins to differ from that of the MS-II due to resolution effects.}
\label{fig:observational}
\end{figure*}

\section{Bimodality in the SMHM Relation}
\label{sec:SMHM_full}

Since we now understand what causes the growth of the quiescent Milky Way-mass galaxy population in the context of this model, we can discuss how this might affect the entire central galaxy population. In Figure~\ref{fig:MWEvol_all}, we showed the evolution of the SMHM relation between  $z = 2.07$ and $z = 0$. In Figure~\ref{fig:redblue_Evoldist}, we show the evolution of the SMHM relation since $z = 2.07$ where we have split 0.2\% of the whole central galaxy population into star-forming (blue dots) and quiescent (red dots) at each redshift. 

Much like what we saw in Section~\ref{sec:SFQPopulations} for the Milky Way-mass galaxy population, the majority of galaxies at $z = 2.07$ are star-forming with a gradual increase in the number of quiescent galaxies at each subsequent redshift. We can also see a familiar correlation between higher mass haloes and earlier quenching times in the whole central population. In fact, in this model, the high mass end of the SMHM relation seems to form solely as a result of quenching. If we look at only star-forming galaxies, we see a single power-law distribution with scatter. The quenched population evolves off to the right of the star-forming `main sequence' of the SMHM relation. Although these two populations are fairly distinct, there is not a specific halo mass at which central galaxies are quenched. This reflects the importance of black hole mass in quenching. In lower mass haloes where the rapid infall regime of cooling operates, we expect halo mass growth to play a more important role in quenching since the slope of the heating-cooling equilibrium boundary causes quenching to be a stronger function of halo mass (See Section~\ref{sec:quenching}). 

In addition, the emergence of a shallow distribution at the high halo mass end of the SMHM relation at late times as a result of quiescence has important implications for studies attempting to link galaxy stellar masses to their host halo masses. In H15, the star formation efficiency of galaxies as a function of halo mass increases until these galaxies become quiescent, accounting for a peak in a plot showing the stellar mass--halo mass ratio versus halo mass (See the lower left panel of Figure~\ref{fig:MWEvol_all}). Quiescence causes $M_{*}/M_{\rm{h}}$ to decrease after a certain point because stellar mass growth via star formation stops while halo mass growth continues.This differs from previous studies that model an evolving SMHM relation with a smoothly varying star formation efficiency  \citep[e.g.][]{bwc2013, mnw2013}. We discuss the implications of this difference in greater detail in Section~\ref{sec:discussion}.

\section{Observational Signatures of Quiescence}
\label{sec:observational}

In an attempt to elucidate what physical mechanisms are behind quiescence, a number of efforts have explored trends in the fraction of quenched galaxies as a function of galaxy parameters using observational datasets at $z < 2$ \citep{swm2007, bvp2012, bep2014}. As in this work, some have attempted to focus on central galaxies, either explicitly \citep[e.g.][]{b2008}, or by noting that at Milky Way masses and above, most galaxies are centrals in their own haloes \citep[e.g.][]{wfv2011}. Restricting our attention to parameters directly inferred from observations, the fraction of quenched galaxies appears to vary with stellar mass, stellar surface density within the half light radius or 1 kpc, inferred velocity dispersion ($\propto$ M/R), Sersic index, and bulge to total mass (B/T) ratio \citep{khw2003, bwm2004, fvs2008, b2008, cfk2012, bvp2012, lws2014, bme2014}. The broad consensus is that the degree of bulge domination appears to be the parameter with which quiescence varies the most strongly \citep[e.g.][]{bme2014, lws2014}. \citet{lws2014} explicitly compare with the \citet[with developments by \citealt{psp2014}]{shc2008, sgp2012} semi-analytic models, arguing that the observed strength of the correlations of quiescence with B/T ratio could come from a strong dependence of quiescence on the black hole mass, with galaxies that have more massive black holes being substantially more likely to be quiescent. In fact, \citet{bep2014} found a correlation between the quiescent fraction and their black hole mass estimate, inferred by the joint consideration of bulge mass and velocity dispersion.

In Figure~\ref{fig:observational} we present visualizations of the quenching behavior in H15 for central galaxies as a function of stellar mass (left), halo mass (middle), and black hole mass (right). The top panels show the distributions of 0.2\% of all central galaxies where star-forming galaxies are blue dots and quiescent galaxies are red dots. The bottom panels show the same axes but we plot the distributions for 2.5\% of all main progenitors of Milky Way-mass galaxies in colored dots over 0.2\% of the whole central population in gray dots. In these panels, as before, blue dots represent galaxies that have remained star-forming since $z = 2.07$, green dots represent galaxies that are star-forming at the given redshift but will become quiescent by the present day, and red dots represent galaxies that have quenched and that will remain quiescent up to the present day. 

In the top panels we note the growth of the quiescent population as a function of redshift which was already seen in Figure~\ref{fig:Evoltracks_rb} and~\ref{fig:redblue_Evoldist}. In accord with observations, the models show a broad and scattered correlation between quiescence and stellar mass, where at higher stellar masses the population is more quenched. Also in accord with observations is the wide range of star formation rates, centred largely on the mass of the Milky Way, where one has both quenched and star-forming central galaxies \citep[e.g.][]{khw2003}. 

In the centre panels one can see that quiescent galaxies tend towards higher halo masses, a trend that was already seen in Figure~\ref{fig:Evoltracks_rb}. An important implication here is that, at least in the context of this model, quenched galaxies with similar stellar masses as star-forming galaxies are likely to live in a substantially more massive dark matter haloes. This leads to the expectation that quenched galaxies at a given stellar mass have a considerably increased number and wider velocity distribution of satellite galaxies, a higher average weak lensing signal, a higher incidence of bright satellites/companions, and if globular cluster number scales with halo mass, a larger number of globular clusters. Many of these expectations are in qualitative accord with weak lensing measurements \citep{vvh2014, mwz2015} and globular cluster number and specific frequency \citep{hhh2014}. In fact, \citet{ww2012} found that central galaxies with stellar masses larger than that of the Milky Way have a significantly larger number of satellites if the central is red in color, or quiescent, than if a galaxy of the same stellar mass is blue, or star-forming. These observations strongly suggest that quenched centrals do in fact live within larger dark matter haloes.

The upper and lower sets of plots at the far right show the relationship between SFR and black hole mass, showing that quiescent galaxies are expected to have more massive black holes than star-forming galaxies, as is expected from our analysis in Section~\ref{sec:quenching}. We note that there is still some scatter between quiescence and black hole mass in the upper panel, with a tail of quenched galaxies having low black hole masses. This emphasizes one of the main messages of Figure~\ref{fig:MbhMhalo_hc} -- that, in this model at least and likely in the universe, quiescence is a function of a number of physical parameters and joint consideration of two or more variables is likely to be important for illuminating the causes of quiescence.

In the lower panel, we present Milky Way-mass galaxies which have a narrow range in stellar mass but a large range of black hole and halo masses. For this range of halo masses, quenching is almost uniquely a function of black hole mass and there is a very narrow region of black hole masses where a galaxy's SFR plummets down to low values. This characteristic value of black hole mass reflects the shallow equilibrium boundary for galaxies in the cooling flow regime, as we explained in Section~\ref{sec:BHmassdependence}. The central galaxy population as a whole does not exhibit this behavior because it samples a much wider range of halo masses which include cooling via both the rapid infall and cooling flow regimes. While this provides a physical mechanism for quenching in this model, it is not clear whether this behavior is reflected in the observations. 

\section{Discussion}
\label{sec:discussion} 

The goals of our study were (1) to understand the parameters important in quenching galaxies and (2) to analyse the way quiescence affects scatter in the growth histories of central galaxies with the stellar mass of the Milky Way. To summarize our results, we began by showing that the dominant quenching mechanism in H15 is AGN feedback, a process that primarily depends on halo mass and black hole mass. We also found that Milky Way-mass galaxies at the present day have an increasingly large range of progenitor stellar masses towards higher redshifts. This diversity in growth history is correlated with present-day halo mass, where more massive haloes tend to have built up their stars earlier than lower mass haloes. When splitting the Milky Way-mass galaxy population into star-forming and quiescent galaxies, we found that present-day quenched galaxies underwent a staggered quenching, where their present-day halo mass correlates with the redshift at which they quench. Finally, we showed that while halo mass is a useful parameter with which to characterize quenching, black hole mass is a much better indicator of quiescence at a specific redshift.

We are cognizant that our analysis is likely to be affected by simplifications and model choices made in the course of developing the H15 semi-analytic model. Our results are likely to be rather sensitive to the time-scale for the reincorporation of gas into the hot halo in the H15 model (See Section~\ref{sec:Henriques}). This particular part of the model is important in determining when and how much hot gas can cool back onto the galaxy. This effect causes gas to be reincorporated into lower mass systems at later times so that the galaxy stays star-forming for a longer time period. Larger systems reincorporate gas on much shorter time-scales which causes them to build most of their stellar mass at early times and cease star formation earlier. This change in the reincorporation recipe is particularly important in better matching the stellar mass function at intermediate redshifts (comparing H15 with e.g., \citealp{gwb2011}, see also \citealp{wpo2012} for a more in-depth discussion of the issue). It is currently unclear if this simplified prescription for the reincorporation of gas is the most robust or physically-motivated way to delay the consumption of cold gas in low mass galaxies at intermediate and high redshift. 

We also note that the internal structure of the gas disc and the stellar distribution, and their fates upon merging are modeled by H15 using highly simplified prescriptions. Choices about how to calculate the sizes, assign characteristic velocities, and estimate SFRs vary from model to model \citep{kpt2015, pg2014, cbn2013}, and in the future it would be useful to explore the importance of such choices on the inferred diversity of growth histories of Milky Way-mass galaxies.

Perhaps more importantly, the prescriptions for quenching are also highly simplified, depending primarily on the balance between the cooling rate of the hot halo gas and the heating rate of the energy input from AGN feedback, which in turn depends on the black hole mass and hot halo mass. This results in a particularly clean separation between quiescent and star-forming galaxies as a joint function of primarily black hole mass and secondarily halo mass at a given redshift. While we acknowledge that this picture is highly simplified, we also point out that it can and should be observationally tested. A census of star formation activity in central galaxies with reliable black hole and halo mass measurements may help characterize the heating-cooling equilibrium boundary, if it indeed exists, giving insight into how halo mass and black hole mass play into the quenching of the central galaxy. Unfortunately, halo mass measurements are not likely to be available for individual galaxies at this mass range in the near future and, as a result, reliable proxies for characterizing the depth of the potential well would need to be determined.

Since we show in Sections~\ref{sec:BHmassdependence} and~\ref{sec:observational} that black hole mass correlates well with quenching, we can also posit that the way in which black holes grow at the centre of their host central galaxies is extremely important in affecting the galaxy's future stellar mass growth. Although this growth process is fully understood in this particular model as being dependent on the merger history of galaxies coupled with the amount of cold gas available for consumption (See Section~\ref{sec:BHmassdependence}), how and when, in detail, black holes grow in the real universe is not fully understood. Given the importance of AGN feedback to Milky Way-mass galaxy growth, we caution that a more complete understanding of black hole growth is necessary in order to better model this feedback.

With regards to identifying progenitors of Milky Way-mass galaxies, previous studies have focused on determining their median stellar mass evolution in order to attempt to observationally understand how these galaxies grow \citep{vwb2010, pff2011, vln2013, pvf2013}. While we do match the median mass growth from progenitor studies that agree with the observed stellar mass function and take into account merging and other effects, very few of these have taken into account the intrinsic scatter of progenitor galaxy stellar masses characteristic of MW-mass galaxy growth in models of galaxy formation. This introduces difficulties when observationally identifying the progenitors of galaxies of a specific present-day stellar mass. Since progenitors of star-forming and quiescent galaxies are likely to be systematically different, this must be correctly accounted for when identifying progenitor populations.

Our results also point to a potentially important limitation of the abundance matching technique for linking galaxies and dark matter haloes \citep{vo2004, kbw2004, cwk2006, cw2009, gwl2010, tkp2011, bwc2013}. These models explicitly match galaxies to haloes by assuming these two properties are monotonically correlated apart from some purely statistical scatter. While this model is fairly simple and agrees with clustering measurements, the basic assumptions it relies upon are inherently uncertain due to our lack of understanding with regards to galaxy growth within haloes. By construction, halo masses do not depend on any attributes of the galaxy other than its stellar mass in most of these models, which is in strong contrast to the physical prescriptions used in semi-analytic modeling. For example, many of these studies do not differentiate between star-forming and quiescent galaxies, a necessary distinction in order to account for the emergence of the flat distribution at high halo masses in the SMHM relation according to our study (See Section~\ref{sec:SMHM_full}). The necessity of this differentiation is also implied by observational studies which suggest quiescence correlates with halo mass estimates from satellite abundances and gravitational lensing \citep{ww2012, mwz2015}.

While the H15 model simulates a population that broadly follows a double power-law fitting of the SMHM relation -- a fit that many others have used before -- the scatter in the relation is substantial and worth noting explicitly. Such scatter has been incorporated in recent generations of models. For example, \citet{bwc2013} assumes a scatter in stellar mass given a halo mass following a lognormal distribution. The fact that the scatter in stellar mass at fixed halo mass is a weak function of halo mass means that abundance matching works relatively well at reproducing this scatter even with the potentially invalid assumption of a lognormal scatter \citep{tkw2004, gcf2015}, as is seen in Figure~\ref{fig:MWEvol_all}. Even so, in the H15 model, where the scatter arises from the astrophysical modeling, we find it to be asymmetric, with long tails towards high halo masses for Milky Way-mass galaxies.

We saw in Sections~\ref{sec:SFQPopulations} and~\ref{sec:observational} that not only is there a significant amount of scatter in the SMHM relation in this model, but that this scatter strongly correlates with galactic properties, such as those that are important for quenching. For example, the scatter in halo masses and growth histories systematically correlates with star formation activity, such that quiescent galaxies typically live in higher mass haloes with a variety of quenching times. This has important implications for studies attempting to explore star formation activity and histories in a halo framework \citep[e.g.][]{hwv2014}. For example, the opposite trend is predicted by the recent age-matching models of \citet[][see \citealt{mwz2015}]{whb2015} which is likely to be a serious limitation to using them to study galaxy evolution. While we acknowledge that semi-analytic models are simplified, they include prescriptions for a diversity of physical processes, and this model in particular agrees with both the observed stellar mass functions and the observed fractions of quiescent and star-forming galaxies over the range of redshifts and stellar masses relevant for this study. Accordingly, this model is a reasonable \textit{qualitative} guide to how the real universe might differ from the assumptions underlying abundance matching analyses. As a result, studies of galaxy growth, especially when focusing on specific galaxy populations, should be aware of these caveats before relying on abundance or age matching techniques. 

\section{Conclusions}
\label{sec:conclusions}

Galaxies appear to be particularly diverse at stellar mass scales similar to that of the Milky Way, where bulgeless star-forming disc galaxies coexist with centrally-concentrated quiescent galaxies. Our goal in this study was to use the semi-analytic model developed by \citet{hwt2015} to explore the diversity of growth histories of central galaxies with stellar masses similar to that of the Milky Way, focusing particularly on how the quenching of star formation affects their growth histories.

The growth history and quenching of central galaxies in this model correlates jointly with black hole mass and halo mass, where quiescent galaxies are those in which AGN heating exceeds the halo cooling rate (Section~\ref{sec:quenching}). This results in a scattered relation between stellar mass and halo mass for central galaxies (Figure~\ref{fig:MWEvol_all}). While this scatter is quantitatively similar to values found in previous studies \citep[e.g.][]{bwc2013, mnw2013}, it is strongly correlated with the physical properties of the central galaxies, a fact that has often been overlooked despite its important implications for the diversity of galaxy growth histories. Central Milky Way-mass galaxies in H15 show a wide diversity in growth histories, from galaxies that constantly form stars since $z \sim 2$ to quenched galaxies which have very little $z < 2$ star formation. 

We find that the quenching of star formation is a significant source of scatter in central galaxy growth histories in H15 (Figure~\ref{fig:Evoltracks_rb}) since it causes the stellar mass buildup of some galaxies to significantly slow down. More specifically, the time at which these galaxies quench correlates with present-day halo mass (Figure~\ref{fig:staggeredquenching}) -- a phenomenon we call staggered quenching. This creates a link between quenching and halo mass, where, at a fixed stellar mass, more massive haloes at the present day tend to have quenched earlier than lower mass haloes. While halo mass correlates better with quiescence than stellar mass, we found that the central black hole mass correlates best with the quenching of galaxies of this stellar mass (Figure~\ref{fig:observational}). While many galaxies experience rapid black hole growth via merging prior to quenching, there do exist systems which become quiescent more gradually. In these systems, while the AGN heating rate is constant, gas cooling becomes less effective at low redshifts. This can stop star formation in galaxies close to the heating-cooling equilibrium boundary (Section~\ref{sec:BHmassdependence}). In addition, the H15 model shows a pronounced heating-cooling equilibrium boundary driven by AGN feedback. This is an observationally testable prediction of a boundary in the $M_{\rm{BH}}$-$M_{\rm{h}}$ relation and future work should focus on searching for it. 

Our results are also important for attempts at observationally identifying progenitors of Milky Way-mass galaxies. Our description of ``staggered quenching'' points to a correlation between halo mass and quenching time that, if at least indirectly observed in the real universe, may give clues to a galaxy's growth history. More importantly, perhaps, is a more complete understanding of the connection between quiescence and the mass of a galaxy's supermassive black hole. Our analysis showed that black hole mass is a better predictor of quiescence than halo mass in the H15 model, so understanding how black holes grow in the real universe seems to be an important factor to understanding the growth of progenitors of quenched galaxies. In terms of other models, abundance matching, age matching, and halo occupancy distribution models all make simplifying assumptions in their implementation that are in partial disagreement with our physically-motivated framework. Understanding in detail how quiescence is reflected in the SMHM relation in these models is essential to address the main question of our paper in the context of other frameworks. Our results serve to show the importance of understanding the astrophysics underlying the quenching of star formation in galaxies.

\section*{Acknowledgments}

B.A.T. is supported by the National Science Foundation Graduate Research Fellowship under Grant No. DGE 1256260. The work of BH and SW was supported by Advanced Grant 246797 ``GALFORMOD'' from the European Research Council. This work is partially supported by {\it HST} grant GO-12060. Support for Program number GO-12060 was provided by NASA through a grant from the Space Telescope Science Institute, which is operated by the Association of Universities for Research in Astronomy, Incorporated, under NASA contract NAS5-26555. This publication also made use of NASA's Astrophysics Data System Bibliographic Services. The Millennium Simulation databases used in this paper and the web application providing online access to them were constructed as part of the activities of the German Astrophysical Virtual Observatory \citep[GAVO,][]{lv2006}. We appreciate fruitful conversations with Brian O'Shea, Keren Sharon, Peter Behroozi, and Casey Papovich.

\bibliography{paper.bib}

\bsp

\label{lastpage}

\end{document}